\documentclass[twocolumn]{aastex63}  
\usepackage{natbib}
\usepackage{graphicx}
\usepackage{txfonts}
\newcommand{\BCPA}{\texttt{RunA}} 
\newcommand{\FBFH}{\texttt{RunB}} 
\newcommand{\PFAA}{\texttt{RunC}} 
\newcommand{\PFAB}{\texttt{RunC$^\mathtt{axi}$}} 
\newcommand{\msol}{M$_{\sun}$}

\newcommand{\Omegap}{$\Omega_\mathrm{p}$}
\newcommand{\omegaunits}{km$\,$s$^{-1}$\,kpc$^{-1}$}
\newcommand{\rlun}{$L_{1,2}$}
\newcommand{\rlquatre}{$L_{4,5}$}

\newcommand{\DLz}{$\Delta L_\mathrm{z}$}
\newcommand{\Lz}{$L_\mathrm{z}$}
\newcommand{\Lzunit}{kpc\,km\,s$^{-1}$}

\newcommand{\Eunit}{kpc$^2$\,Myr$^{-2}$}

\newcommand{\dem}{$D_2(E)$}
\newcommand{\dlzm}{$D_2(L_\mathrm{z})$}

\newcommand{\dlzmunit}{(kpc$^2$\,km$^2$\,s$^{-2}$)\,Myr$^{-1}$}
\newcommand{\demunit} {(kpc$^4$\,Myr$^{-4}$)\,Myr$^{-1}$}

\renewcommand{\vec}[1]{\mathbf{#1}}

\received{August 28, 2019}
\revised{November 19, 2019}
\accepted{December 5, 2019}
\submitjournal{ApJ}

\reportnum{astro-ph/XXX}

\shorttitle{Stellar migration in galaxy discs using the Chirikov diffusion rate}
\shortauthors{Wozniak H.}

\begin{document} 

\title{Stellar migration in galaxy discs using the Chirikov diffusion rate}
\correspondingauthor{Herv\'e Wozniak}
\email{herve.wozniak@umontpellier.fr}

   \author[0000-0001-5691-247X]{Herv\'e Wozniak}

   \affiliation{LUPM, Univ Montpellier, CNRS, Montpellier, France}

   \begin{abstract}
 
We are re-examining the problem of stellar migration in disc galaxies
from a diffusion perspective.
We use for the first time the formulation of the diffusion rates
introduced by \citet{1979PhR....52..263C}, applied to both energy $E$
and angular momentum \Lz\ in self-consistent N$-$body experiments. We
limit our study to the evolution of stellar discs well after the
formation of the bar, in a regime of adiabatic evolution.
We show that distribution functions of Chirikov diffusion rates have
similar shapes regardless the simulations, but different slopes for
energy and angular momentum. Distribution functions of derived
diffusion time scales $T_D$ have also the same form for all
simulations, but are different for $T_D(E)$ and
$T_D(L_\mathrm{z})$. Diffusion time scales are strongly dependent on
\Lz.  $T_D(E) \lesssim 1$~Gyr in a \Lz\ range roughly delimited by the set
of stellar bar resonances (between the Ultra Harmonic Resonance and
the Outer Lindblad Resonance). Only particles with low \Lz\ have
$T_D(L_\mathrm{z}) \lesssim 10$~Gyr, i.e. the simulation length. In terms
of mass fraction, 35 to 42\%\ turn out to diffuse energy in a
characteristic time scale shorter than 10~Gyr, i.e. simulations
length, while 60 to 64\% undergo the diffusion of the angular momentum
on the same time scale.
Both the diffusion of \Lz\ and $E$ are important in order to grasp the
full characterisation of the radial migration process, and we showed
that depending on the spatial region considered, one or the other of
the two diffusions dominates.

    \end{abstract}


%

\section{Introduction}
\label{sec:introduction}

Stellar migration of the galactic disc stars has been invoked as a
dynamical mechanism to explain the dispersion of stellar metallicity
observed in the solar neighbourhood. The age$-$metallicity relation
(AMR) shows that the dispersion of stellar metallicity increases with
the age of the stars
\citep[e.g.][]{1993A&A...275..101E,2008MNRAS.388.1175H,2008A&A...480...91S,
  2019A&A...624A..19B}. Another relation, the age$-$velocity
dispersion relation (AVR)
\citep[e.g.][]{1977A&A....60..263W,2008A&A...480...91S,2019MNRAS.489..176M}
suggests the existence of a heating mechanism of the stellar
disc. Stellar migration could then also be a cause, although other
mechanisms have been proposed.

However, stellar migration does not take place in a hypothetical
perfectly axisymmetric disc, made of stars rotating in circular
orbits. One or more gravitational perturbations are at the origin of
any deviation from this hypothetical perfection. These perturbations
can be intrinsic (density waves such as bars or spiral arms, two-body
relaxation, ...) or extrinsic (galaxy satellites, encounters, mergers,
gas accretion...). Also, the amplitude of the perturbations, and thus
their ability to reproduce observations, depends on the underlying
physical mechanism invoked.

Two classes of models have recently been proposed. They have been the
focus of attention since then. All of them distinguish between the
effects of blurring (i.e. the radial migration of a star is due to
epicyclic motion around a fixed guiding radius), and churning
(i.e. the radial migration is due to a change in this guiding radius).

\citet[][]{2002MNRAS.336..785S} have shown that spiral waves, possibly
transient, have the ability to modify the angular momentum of stars
without changing the distribution function, so that the disc does not
heat up as a result of these changes.  These angular momentum changes
essentially result in a variation in the mean radius of stellar orbits
over time while keeping their low eccentricity. The dominant
  mechanism is thus churning. These spiral waves have their own
pattern speeds with which stars may
resonate. \citet[][]{2002MNRAS.336..785S}, confirmed by
\citet[][]{2012MNRAS.426.2089R}, have shown that angular momentum
exchanges take place mainly at corotation. Therefore, the corotation
scattering mechanism might be responsible for stellar migration.

For \citet[][]{2010ApJ...722..112M}, resonances are also responsible
for stellar migration, but their mechanism differs somewhat from
\citet[][]{2012MNRAS.426.2089R}. Indeed,
\citet[][]{2010ApJ...722..112M}, confirmed by
\citet[][]{2011A&A...527A.147M}, consider the interactions between a
stellar bar and a spiral structure. In this case, at least two
patterns exist and the resonances may overlap. Resonance overlap
introduces additional chaos by increasing the efficiency of orbit
scattering, which also modifies the angular momentum of stars.
Indeed, motion in chaotic regions can be diffusive
\citep[e.g.][]{1983A&A...117...89C}

We are therefore faced with two mechanisms, which are not
irreconcilable from the point of view of galactic dynamics, but which
have different observational consequences on the AMR and AVR. Indeed,
a related important question for the AVR is whether or not these
phenomena cause just the right amount of chaos in the disc to explain
increases in the velocity dispersion over time.  Whether we deal with
stellar migration, disc heating, or stochasticity, stellar motions can
be studied from different viewpoints. Indeed, each analysis uses a
different methodological framework but the fundamental observational
fact is that stars do not stay at their birth site. Therefore, the
fundamental question, which is still under debate, is not ultimately
to know which dynamical process is solely responsible for the radial
migration of stars, but rather what are the relative intensities of
each of these processes, whether they contribute together to the same
phenomenon, or whether they are ultimately only different points of
view of the same phenomenon whose root cause should still be
determined.

We have decided to tackle the problem with tools of non-linear
physics. This article is only a preliminary step towards answering the
fundamental questions mentioned above. Our approach here is to
reanalyse the {\em diffusion} of quantities such as energy and angular
momentum. We have measured diffusion by applying for the first time
the Chirikov diffusion coefficient to galactic N-body simulations.
\citet[][]{2011A&A...534A..75B} have already addressed the issue but
in the general context of Fokker-Planck diffusion.

After some fundamental considerations on the dynamics of a rotating
disc subjected to perturbations (Section~\ref{sec:theory}), we
introduce the Chirikov diffusion rate in Section~\ref{sec:chirikov},
then N$-$body simulations on which we have applied this tool
(Section~\ref{sec:nbody}). Sections \ref{sec:chirikov_results} and
\ref{sec:timescales} are dedicated to the analysis of the results that
will be discussed in
Section~\ref{sec:discussion}. Section~\ref{sec:axi} focuses on an
axisymmetric case for the sake of comparison.

\section{Angular momentum and energy variations}
\label{sec:theory}

The dynamics of a rotating stellar disc forced by a spiral or a bar
(or any other driving force) is well-known and reminded by
\citet[]{2002MNRAS.336..785S} in the context of stellar migration. 
For any galactic dynamical system subjected to one perturbative
frequency, the only integral of motion is Jacobi's integral defined
as:
\begin{equation}
\label{eq:one}
E_\mathrm{J} = E - \Omega_\mathrm{p} \, L_\mathrm{z}
\end{equation}
where $E$ is the classical energy in the non-rotating inertial frame,
\Lz\ is the angular momentum component on the z-axis $\vec{e_z}$
chosen to be conveniently the rotation axis, and assuming that $\vec{\Omega} =
\Omega_p\,\vec{e_z}$, the frequency of the perturbation.  $E$ is the
sum of the kinetic and potential energy in a closed system.
Therefore, since $\Delta E_\mathrm{J} \equiv 0$, any variation of 
$E$ is linearly related to \Lz\ and vice-versa as $\Delta E =
\Omega_\mathrm{p}\,\Delta L_\mathrm{z}$.

From the point of view of Hamiltonian dynamics, the existence of two
pattern speeds in a galactic disc is similar to a physical system with
motions on two different time scales, the so-called slow-fast
systems. The fast system could be the bar which has the greater
pattern speed whereas a spiral structure could be the slow one. The
same situation occurs with the phenomenon of the bar-in-the-bar
\citep{2015A&A...575A...7W}. For such systems, adiabatic invariants
are important dynamical quantities as approximate integral of motion:
on the one hand the motion over long time ranges is almost regular if
several such adiabatic invariants exist. On the other hand,
dissolution of these invariants is one possible mechanism for onset of
chaotic dynamics. Indeed, resonant phenomena in fast motion lead
generally to dynamical chaos and transport in large regions of the
phase space as they destroy adiabatic invariance.

\begin{figure}[htb!]
  \centering
\includegraphics[keepaspectratio,width=\hsize]{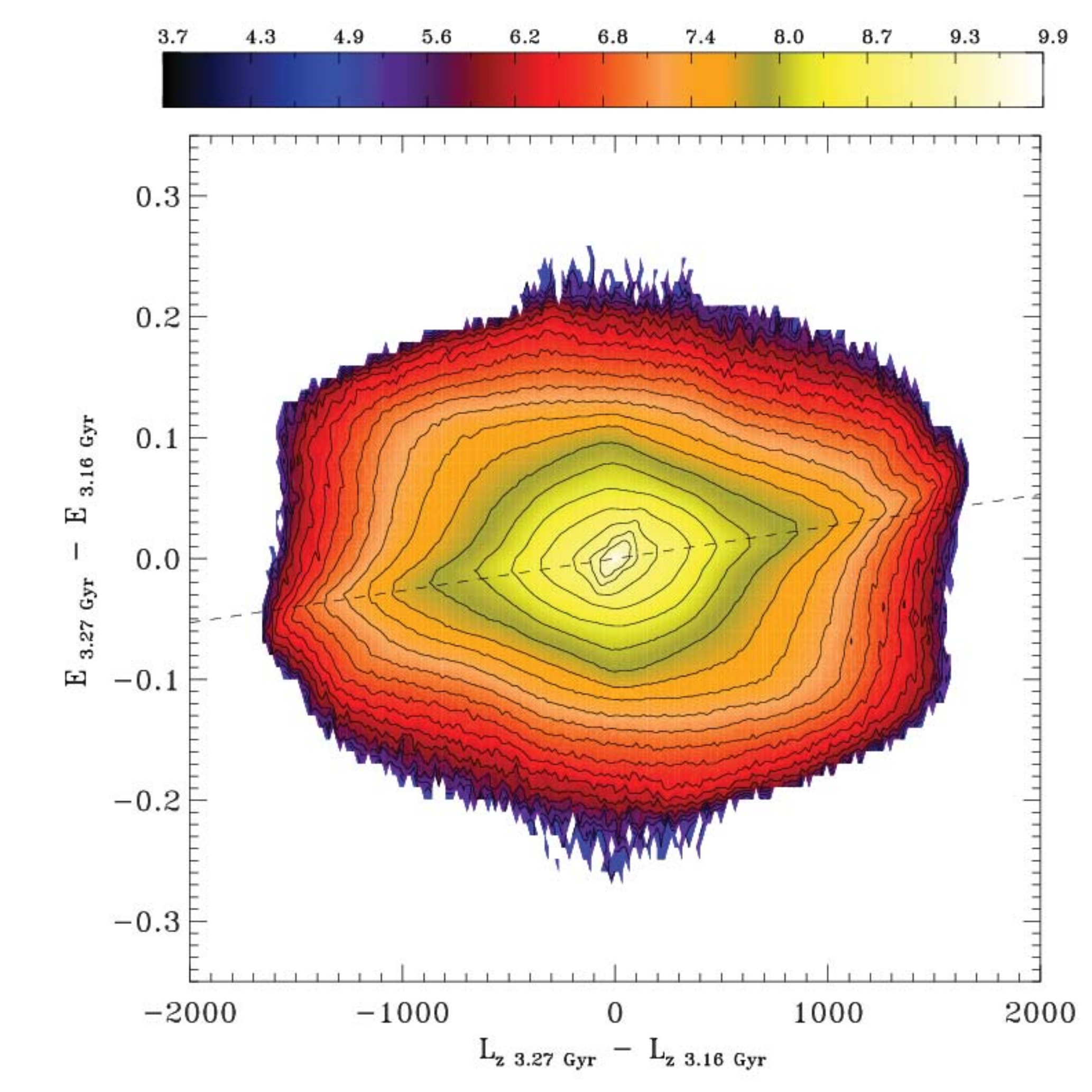}
  \caption{Mass distribution in the plane $\Delta E$ vs \DLz\ for
    \PFAA\ (cf. Section~\ref{sec:nbody}) between times $t\!=\!3.16$
    and $t\!=\!3.27$~Gyr. Colorbar is scaled in log(\msol) per
    bin. The dashed line is the location of $\Delta E =
    \Omega_p\,\Delta L_\mathrm{z}$, where $\Omega_p$ is the bar
    pattern speed.}
  \label{fig:dedlz_pfaa}
\end{figure}

If several patterns coexist in the disc, each one being possibly
variable, the $\Delta E$ vs $\Delta L_\mathrm{z}$ plane must exhibit
several coexisting slopes.  This is what simulations of
Section~\ref{sec:nbody} would suggest. Indeed,
Figure~\ref{fig:dedlz_pfaa} shows that other patterns
exist. Therefore, not all mass transfers have the stellar bar as the
responsible party. This figure shows also that there is a continuum of
$\Delta E$ and $\Delta L_\mathrm{z}$ values for which mass transfers
occurs.

In fact, working with the ratio $\Delta E / \Delta L_\mathrm{z}$ (as
\citet{2002MNRAS.336..785S} and \citet{2012MNRAS.426.2089R} did) masks
too much the complexity of the redistribution in $E$ and \Lz, as this
technique focuses on the dominant patterns, whether they are due to
the bar or the spiral structures. Alternatively, the study of
variations in $E$ and \Lz\ can be based on the difference of these
quantities between two times. These times can be distant: this is the
case between final and initial values, the notion of ``final'' being
understood here as representative of a typical state of the
galaxy. This approach was adopted by all stellar migration studies
since \citet{2002MNRAS.336..785S}, but only for \Lz, with $\Delta
L_\mathrm{z} = L_\mathrm{z} (t) - L_\mathrm{z}(0)$.

$\Delta L_\mathrm{z}(\Delta t) =
L_\mathrm{z}(t_2)-L_\mathrm{z}(t_1)$ is another approach used by, for
instance, \citet{2012MNRAS.426.2089R}. It has the merit of focusing on
the relative variations with respect to an earlier time, whatever the
meaning to be given at that time. The two times can be close, and as
close as we want, so that the difference tends towards a
differential. If the study is limited to the consequences of the
development of certain structures (stellar bar, spiral arms for
example), the initial state can be chosen wisely in order to isolate
the perturbation created by these structures. Finally, pushed at
infinitely small time intervals, this formulation expresses an
instantaneous variation of the angular momentum $\dot{L_\mathrm{z}}$
which is related to the net torque acting on the system of particles.

It should be recalled that, at the level of individual particles,
taking a particular time as a reference situation is not necessarily
more correct than taking the initial time. Indeed, because of the
combined effect of relaxation and bar formation, instantaneous
individual angular momenta may not be representative of time-averaged
angular momenta. For instance, if a particle is able to move {\em
  alternatively} outward and inward in radius while preserving the
circularity of its orbit (what is typical of an epicyclic orbit with
low $\kappa$ frequency), it contributes to the instantaneous $\Delta
L_\mathrm{z}$ taken at any particular time. But averaged on several
rotations, the time-averaged $\overline{L_\mathrm{z}}$ is only
representative of the mean radius, and thus $\Delta
\overline{L_\mathrm{z}} \sim 0$. On the contrary, if that particle
moves adiabatically outward or inward, exclusively, then $\Delta
\overline{L_\mathrm{z}} \neq 0$. The particle then moves to a nearby
region of the phase space, its mean radius and/or rotational velocity
having been modified. There is diffusion.

It is therefore necessary to average the measurements in one
way or another, both on the angular momentum and the energy of
individual particles. The averaging is intended to cancel the
influence of bounded energy/angular momentum oscillations and
emphasise the accumulating changes, related to the diffusion process.

\section{Introduction to Chirikov diffusion rate}
\label{sec:chirikov}

The diffusion of $E$ means that the energy of the body, as a result of
the accumulation of small random variations, can take larger, as well
as smaller values, as compared to the unperturbed energy. Similarly,
diffusion of \Lz\ means that even if the trajectory of unperturbed
motion of the body was close to circular, a perturbation may bring about
trajectories with high eccentricity.  If $E$ increases at
constant rotational velocity, it will necessarily generate an increase
in \Lz\ because the radius increases. But a similar effect can be
achieved by increasing the velocity. Since positions and velocities
vary together, the real diffusion rate depends on both $E$ and \Lz.

In this context, the {\em Chirikov diffusion rate}
\citep[][eq. 4.6]{1979PhR....52..263C} appears to be a natural
choice. Applied on $E$, for individual particles, it is defined as :
\begin{equation}
\label{eq:chirikov}
  D_n(E) = \overline{(\Delta\overline{E})^2/\Delta t}.
\end{equation}
Although the original definition deals only with $E$, we may extend
the definition of Equation~(\ref{eq:chirikov}) to compute
$D_n(L_\mathrm{z})$.  In Equation~(\ref{eq:chirikov}), $\overline{E}$ is the
value of energy averaged over a period of $\Delta t_n=10^n$ (in time
unit of the system). In our case, it is convenient to choose $\Delta
t_n \equiv \Delta t_2$ as the minimum time that separates two
snapshots, i.e. 100 code units (105.49 Myr). Indeed, the whole
simulation (10.54~Gyr) is naturally segmented at regular
intervals. $\overline{E}$ and $\overline{L_\mathrm{z}}$ are thus
computed on the fly for each particles and stored with snapshots.

$\Delta\overline{E}$ is then the difference between two intervals
(snapshots), $\Delta t$ being the time difference between the
snapshots. These snapshots are not necessarily consecutive because the
second averaging concerns all possible pair combinations. This
procedure, initiated by \citet[]{1979PhR....52..263C}, ensures that
all time scales are represented by the definition of $D_n$.

Several other definitions of $D_n$ exist and an abundant literature
concerns the interpretation to be given to the evolution of $D_n$ with
the {\em strength} of the perturbation(s), and its asymptotic
behaviour when $n$ increases \citep[][]{1992rcd..book.....L}.  As a
general rule, $D_n$ sharply increases above a certain threshold of
perturbation strength meaning that the motion is moving from regular
to chaotic. We are not looking here for any critical value of the
perturbation strength, as the definition of this strength can be the
subject of much debate.  Indeed, each region of a galactic disc is
subjected to perturbations of different intensities while all these
regions remain connected through gravitation. It is therefore very
difficult to highlight particular threshold values of the perturbation
intensity in a large N$-$body system, considered as a whole.  Our
objective is rather to qualify the different types of particle
populations with noticeable $D_n$ values, or range of $D_n$, possibly
different from one region to another. This allows to determine whether
some regions are more stochastic than others and, if so, whether $E$
or \Lz\ is the more diffusive quantity.

\section{N$-$body experiments}
\label{sec:nbody}

Several simulations have been performed to check the dependence of our
results on certain quantities, such as the total mass, or the initial
scale of the stellar disc.  Rather than starting from a cosmological
situation, including all kinds of effects that are difficult to
control (accretion of dwarf galaxies, cold gas flows, star formation,
etc.), we preferred to start with an idealised situation.

Initial stellar populations are set up to reproduce such idealised,
but typical, disc galaxies. Positions and velocities for
$N_\mathrm{s}$ particles are drawn from a superposition of two
axisymmetrical \citet{mn75} discs of mass $M_1$ and $M_2$
(cf. Table~\ref{tab:simul}), of scale lengths $l_1$ and $l_2$~kpc and
a scale height $h_z$. Initial positions have been truncated to
$R=30$~kpc for \BCPA\ and \FBFH, $R=40$~kpc for the more massive
\PFAA. Scale lengths and scale heights have been chosen such as the
superposition of the two axisymmetric distributions shapes the initial
spatial configuration as disc galaxy with a small but significant
bulge.

Initial velocity dispersions are computed solving numerically the
Jeans equations according to the \citet{1993ApJS...86..389H} method.
The initial velocity dispersion was chosen to be anisotropic with
$\sigma_r = \sigma_z$ and $\sigma_{\theta}^2 = \sigma_r^2 \kappa^2 /
(4\Omega^2)$, where $\sigma_r$, $\sigma_{\theta}$ and $\sigma_z$ are
three components of the velocity dispersion along respectively the
radial, azimuthal and vertical directions and $\kappa$ and $\Omega$ are
respectively the radial and angular epicyclic frequencies. They are
related by $\kappa^2 = 4\Omega^2+rd\Omega^2/dr$

\begin{figure}[htb!]
  \centering
  \includegraphics[keepaspectratio,width=0.75\hsize]{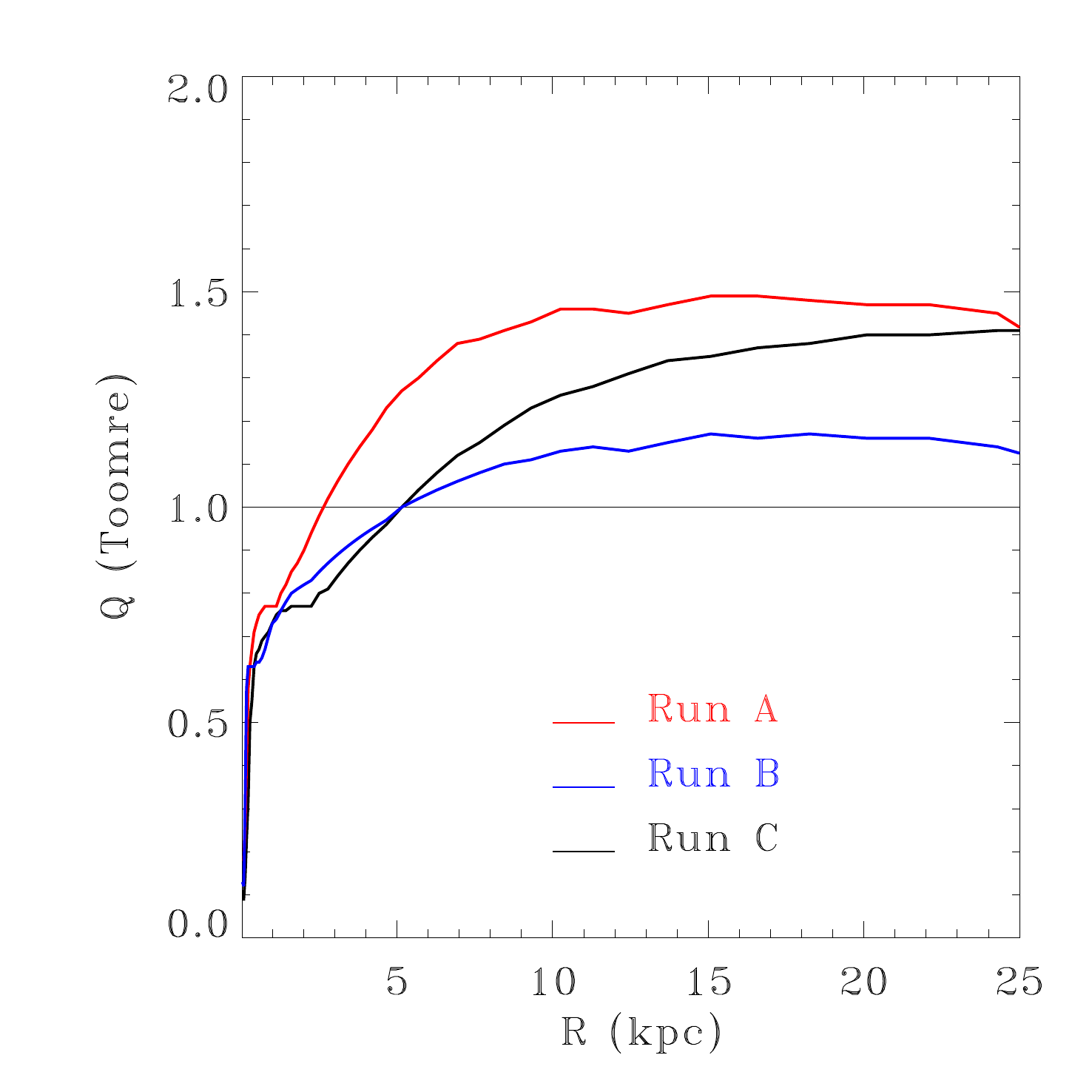}
  \caption{Initial $Q$ Toomre parameter as a function of radius for
    the three simulations.}
  \label{fig:qq}
\end{figure}

As the Toomre parameter ($Q=\sigma_r\,\kappa/(3.36 G \mu)$ where
$\mu$ is the mass surface density and $G$ the gravitational constant)
has not been explicitly constrained, the resulting values are
displayed in Figure~\ref{fig:qq}. The three simulations are unstable ($Q
< 1$) in their central region, i.e. at roughly one scale length around
the centre. This is typically the region where the bar is formed.

\begin{table}[htb!]
\caption{Main initial parameters: name of the run (Model), number of
  stellar ($N_\mathrm{s}$) particles, masses, scale lengths and common
  scale height of the two Miyamoto-Nagai initial distributions ($M_1,
  M_2, l_1=a_1+h_z, l_2=a_2+h_z, h_z$).}
\label{tab:simul}
\flushleft
\begin{tabular}{@{}llllllll@{}}
\tableline 
Model &  $N_\mathrm{s}$ & $M_1$  & $M_2$ &$l_1$ &$l_2$&$h_z$ \cr
      &$\times 10^{7}$&$\times 10^{10}$ M$_{\sun}$&$\times 10^{10}$ M$_{\sun}$&(kpc)&(kpc)&(kpc) \cr
\tableline
\BCPA & 4.          & 0.63    & 3.57  &0.57 &2.0  & 0.5  \cr 
\FBFH & 4.4         & 1.1     & 11.0  &1.0  &3.5  & 0.5  \cr
\PFAA & 4.          & 3.0     & 17.0  &1.14 &4.0  & 1.0  \cr
\tableline
\end{tabular}
\end{table}

\PFAA\ and \BCPA\ use similar initial parameters than respectively
\citet[]{1991A&A...252...75P} and
\citet[]{2011A&A...534A..75B}. \FBFH\ has similar initial conditions
than the run named ``SimS'' in \citet[]{2015A&A...575A...7W} but is
made exclusively of stellar particles for the same total mass. All
runs are computed until 10.54~Gyr.

\begin{figure}[htb!]
  \centering
  \includegraphics[keepaspectratio,width=0.75\hsize]{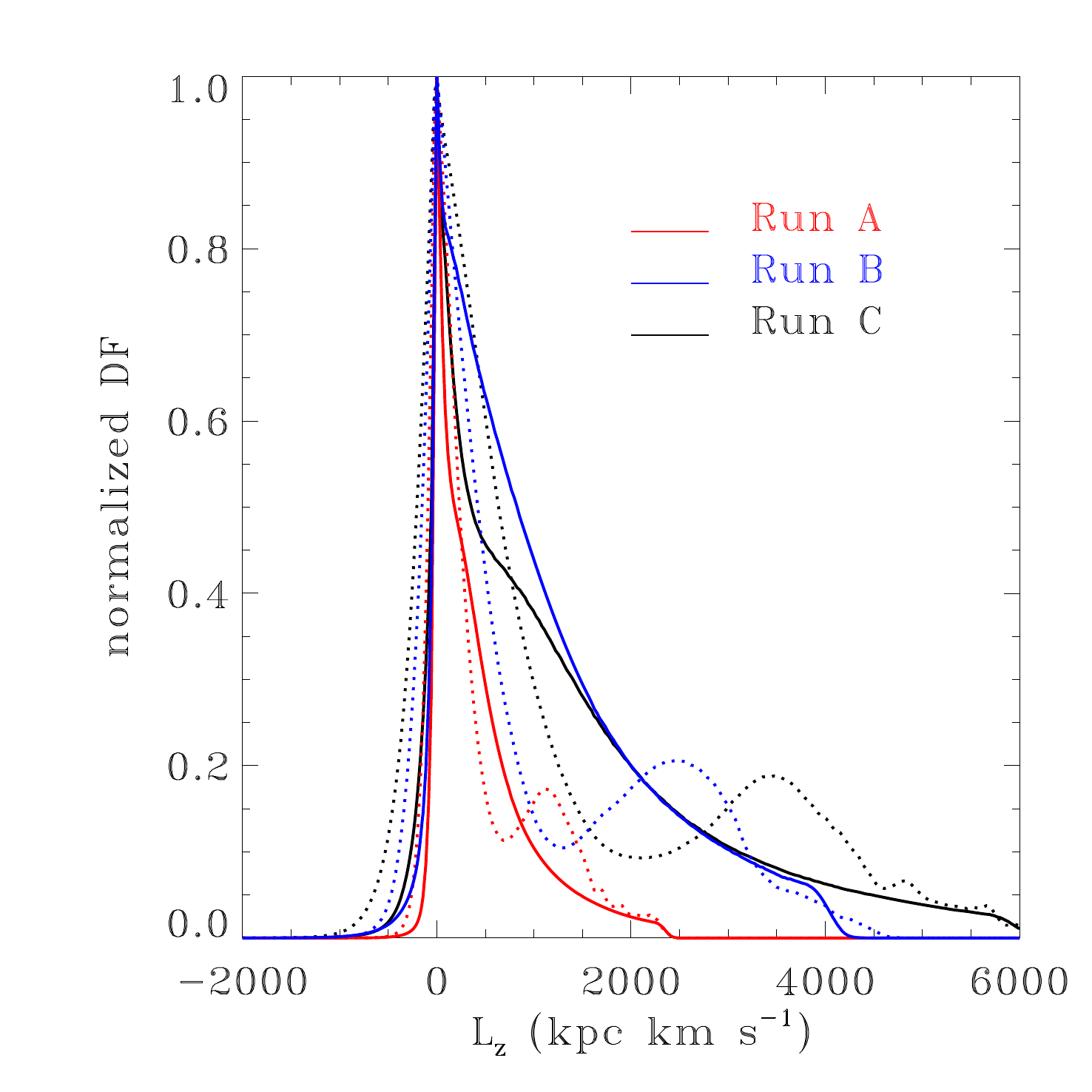}
  \includegraphics[keepaspectratio,width=0.75\hsize]{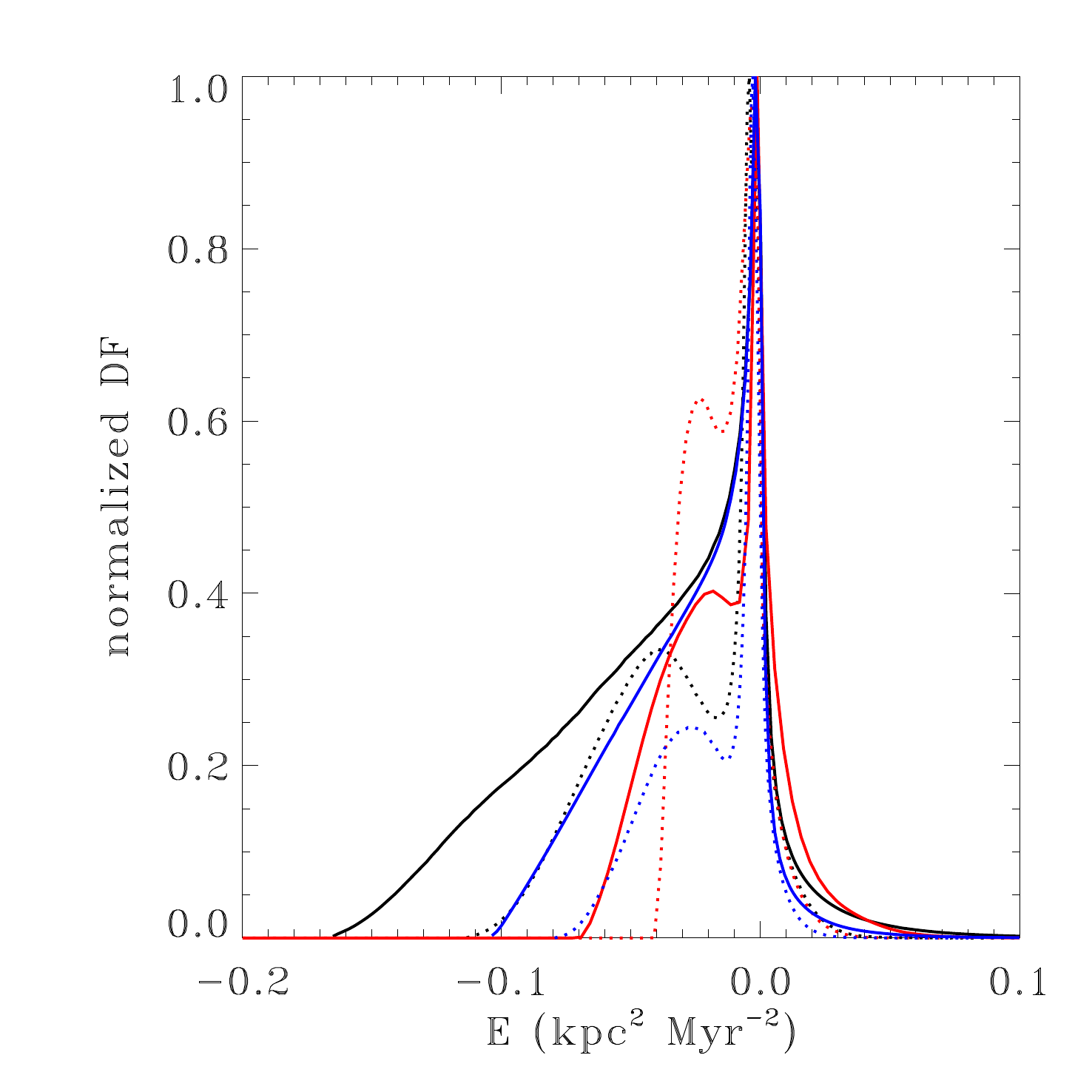}
  \caption{Initial (solid lines) and final (dotted lines) distribution
    functions (DF) for the three simulations. DFs have been normalised
    to DF(0) which is also the maximum. The bump in the final DF,
    around $L_\mathrm{z}\approx$~1100, 2500 and 3500~\Lzunit\ for,
    resp., \BCPA, \FBFH, and \PFAA, is typical of the `hot' particle
    population which is able to cross the corotation and explore both
    the bar and the disc.}
  \label{fig:fdlzinit}
\end{figure}

The evolution is computed with a particle--mesh N$-$body code, derived
from the original version of the Geneva group
\citep{1993A&A...270..561P}.  The broad outline of the code is the
following: the gravitational forces are computed with a particle--mesh
method using a 3D log--polar grid with $(N_R, N_\phi,
N_Z)=(60,64,312)$ active cells. The smallest radial cell in the
central region is 36~pc large and the vertical sampling is 50~pc. The
extent of the mesh is 100~kpc in radius and $\pm 7.8$~kpc in height.
Since we used a polar grid and we need an accurate determination of
the forces in the central region, we have improved the pre-computation
of self-forces by subdividing each cell in $(n_r, n_\phi,
n_z)=(32,6,6)$ subcells. Self-forces are then linearly interpolated
before being subtracted from the gravitational forces.

In a perfectly collisionless simulation of a stable equilibrium model,
each particle would conserve its specific energy.  The combination of
a particle-mesh code, an initial relaxed distribution and a large
number of particles ensures that the sources of numerical diffusion
are minimised. However, we have also performed a control run (\PFAB)
which will be detailed in Section~\ref{sec:axi}.

For convenience, the units in which the discussion will be conducted
have been chosen to avoid the power of 10. Thus, the specific angular
momenta will be in \Lzunit\ while the specific total energies will be
in \Eunit. In addition, since all particles have the same mass, all
particle number distributions can also be read as mass fractions.

The initial and final distribution functions (DF,
cf. Figure~\ref{fig:fdlzinit}) are typical of such type of
simulation. Expressed as a function of \Lz, these DF display similar
trends in regard to the DF obtained by 3D $N$-body simulations,
e.g. those of \citet{1978ApJ...226..521Z}, \citet{1987MNRAS.225..653S}
or \citet{1993A&A...270..561P}.  The shape of these DFs has been
explained by a superposition of various families of orbits
\citep{1997A&A...317...14W, 1999CeMDA..73..149W}. Orbits of the bump
(around $L_\mathrm{z}\approx$~1100, 2500 and 3500~\Lzunit\ for, resp.,
\BCPA, \FBFH, and \PFAA) are mostly disc orbits which also populate
the corotation region of the bar. These orbits spend most of their
time outside the bar and sometimes enter inside the bar from the
$L_{1,2}$ Lagrangian points. This last kind of orbits as well as
Lagrangian orbits form the `hot' population described first by
\citet{1987MNRAS.225..653S}. This `hot' population may contribute up
to 30\% of the total mass.

\begin{figure}[htb!]
  \centering
  \includegraphics[keepaspectratio,width=\hsize]{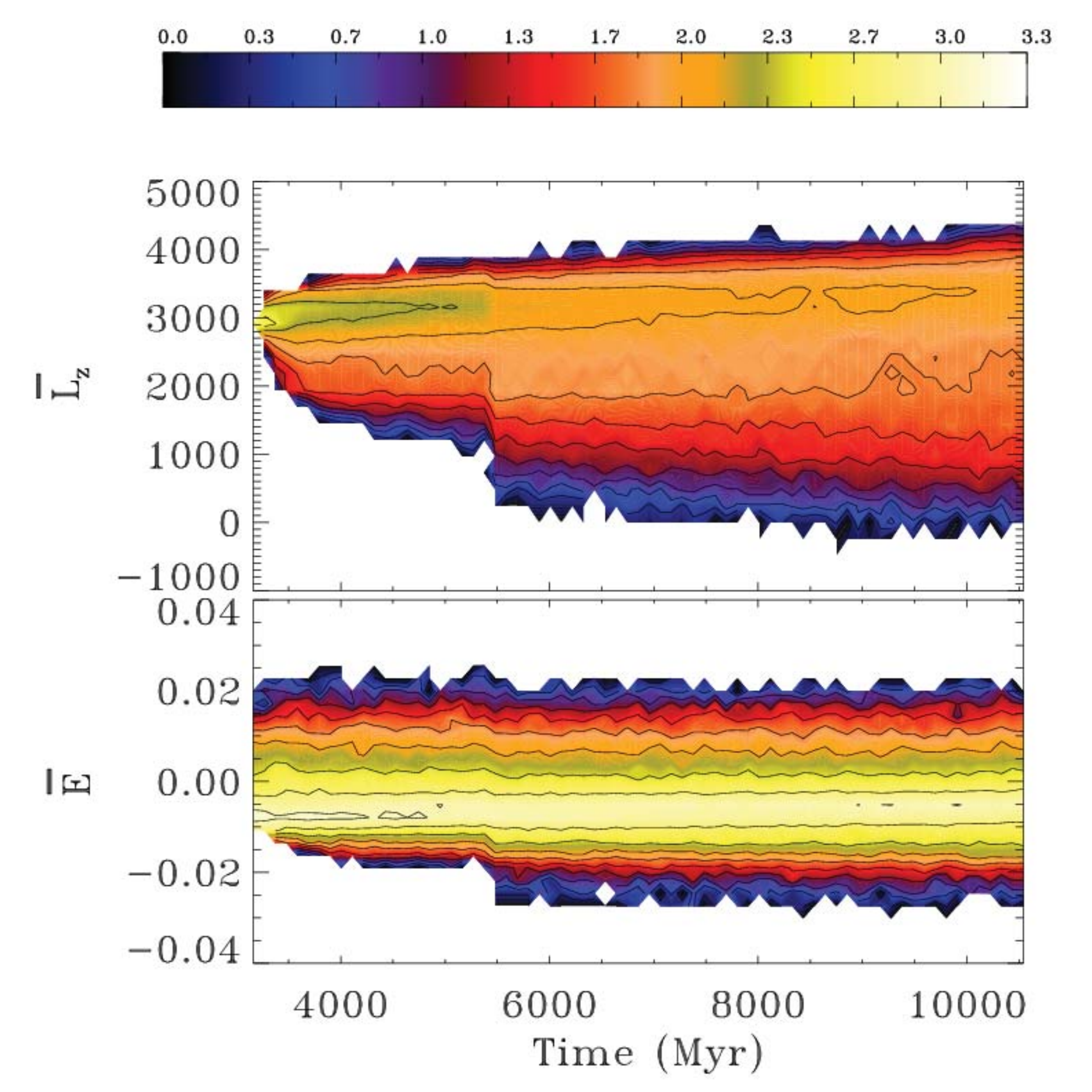}
  \caption{Evolution of $\overline{L_\mathrm{z}}$ (top) and
    $\overline{E}$ (bottom), for 3847 particles selected for
    \PFAA\ (cf. Sect~\ref{sec:nbody} for details) at $t\!=\!3.27$~Gyr
    with $\overline{L_\mathrm{z}} = 3000\pm0.3$~\Lzunit.  Colorbar is
    scaled in log(\msol) per bin.}
  \label{fig:evol}
\end{figure}

Figure~\ref{fig:evol} displays the evolution of
$\overline{L_\mathrm{z}}$ and $\overline{E}$, for a group of particles
selected for \PFAA\ at $t\!=\!3.27$~Gyr with $\overline{L_\mathrm{z}}
= 3000\pm0.3$~\Lzunit, typical of the `hot' population. This selection
represents 3847 particles. Although these particles are selected over
a narrow interval in $\overline{L_\mathrm{z}}$, the values of
$\overline{E}$ show initially a larger amplitude, the maximum being
around $-0.0075$~\Eunit. As the group evolves, the amplitude of
$\overline{L_\mathrm{z}}$ increases rapidly, until it reaches a range
of values from $\approx 0$ (or even negative for some particles) to
$\approx 4500$.  The $\overline{L_\mathrm{z}}$ distribution mode
increases until it reaches a value of $\approx 3500$ at
$t\!=\!10.54$~Gyr. Values of $\overline{E}$ also vary over the same
time interval. However, its distribution gradually spreads only on the
negative side.  Let us recall here that these are average values over
an interval of $\Delta t_2 \!=\!105$~Myr, i.e. between an half and a
quarter of the bar rotation period. Averaging over a larger $\Delta t_n$
does not change the result.

\begin{figure*}[htb!]
  \centering
\includegraphics[keepaspectratio,width=0.33\hsize]{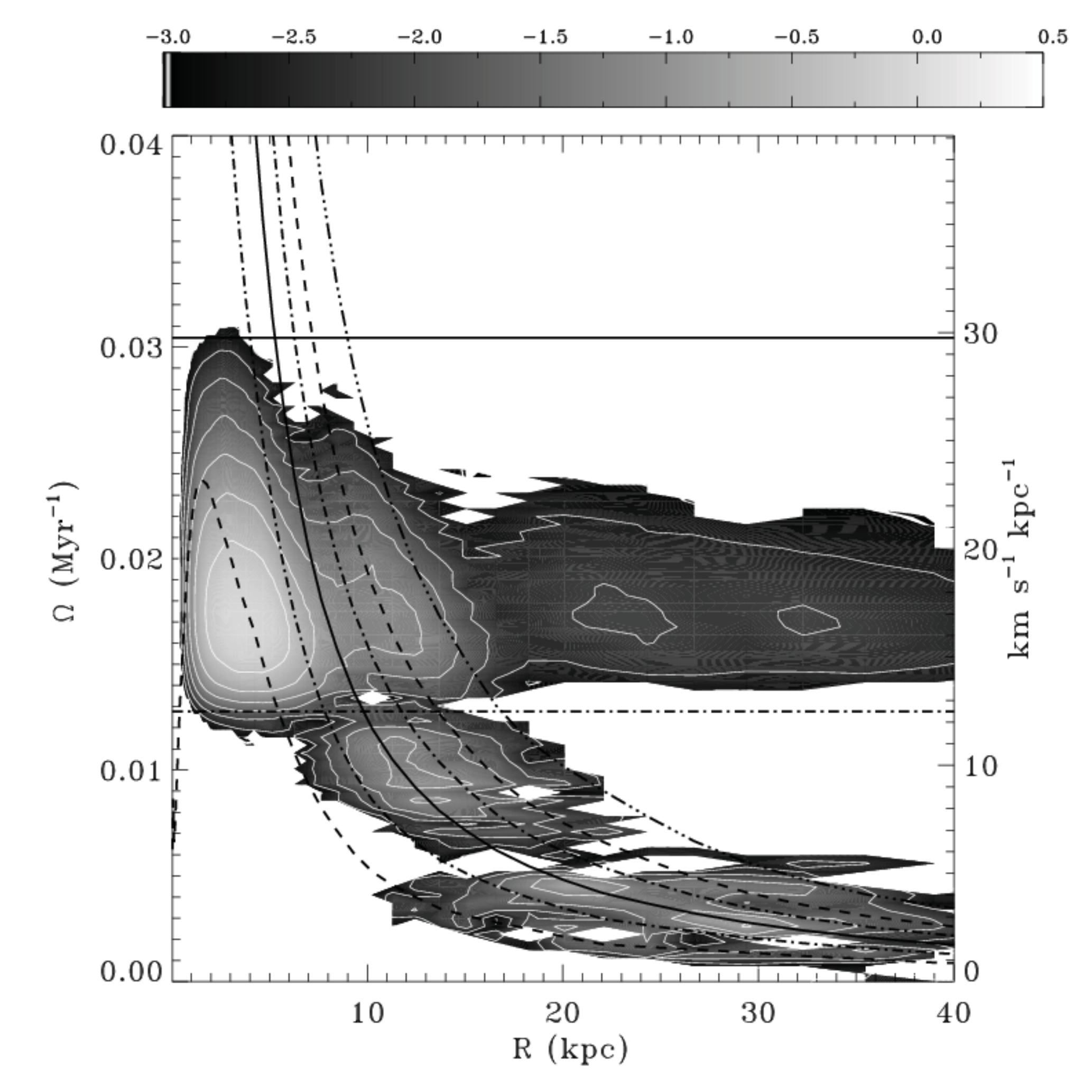}
\includegraphics[keepaspectratio,width=0.33\hsize]{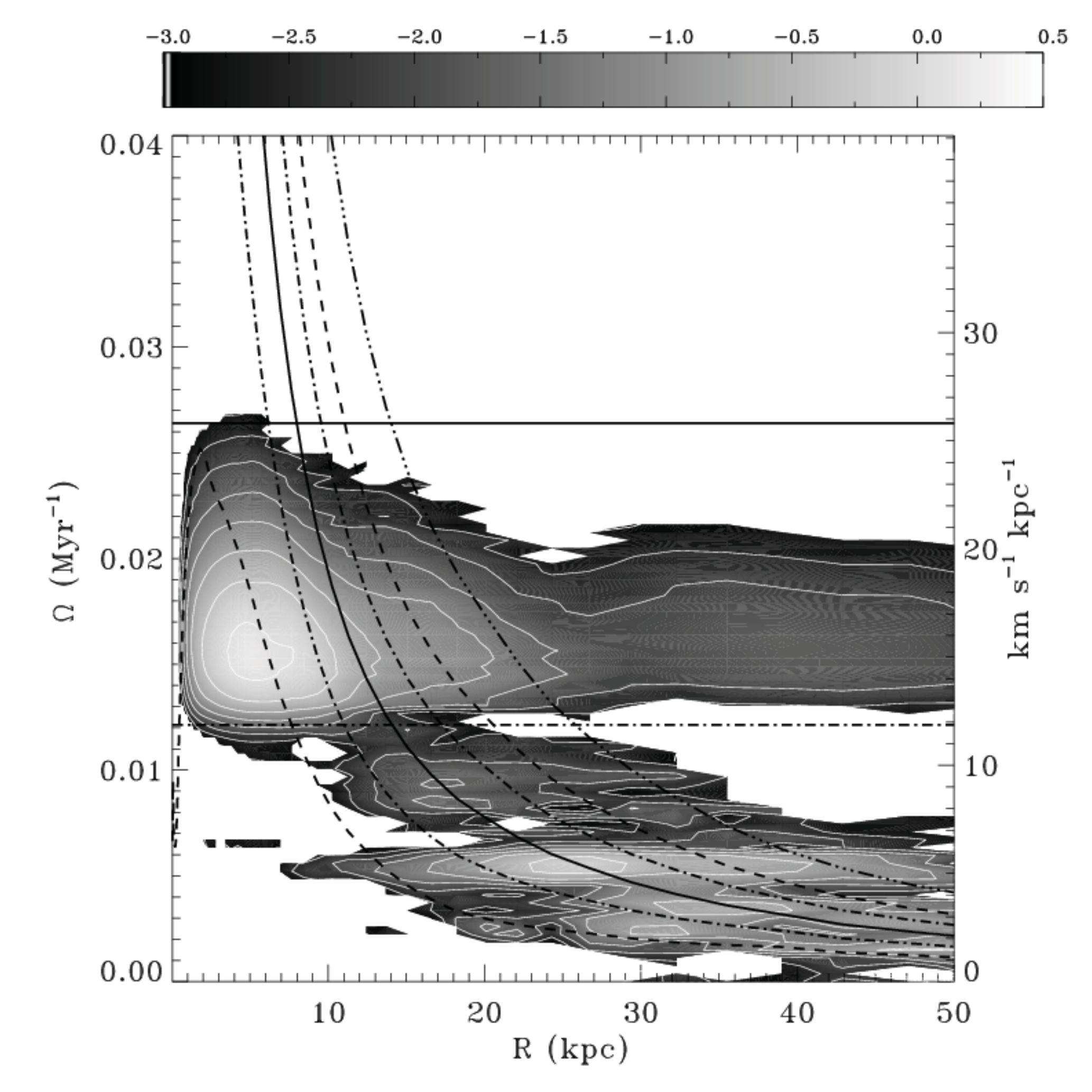}
\includegraphics[keepaspectratio,width=0.33\hsize]{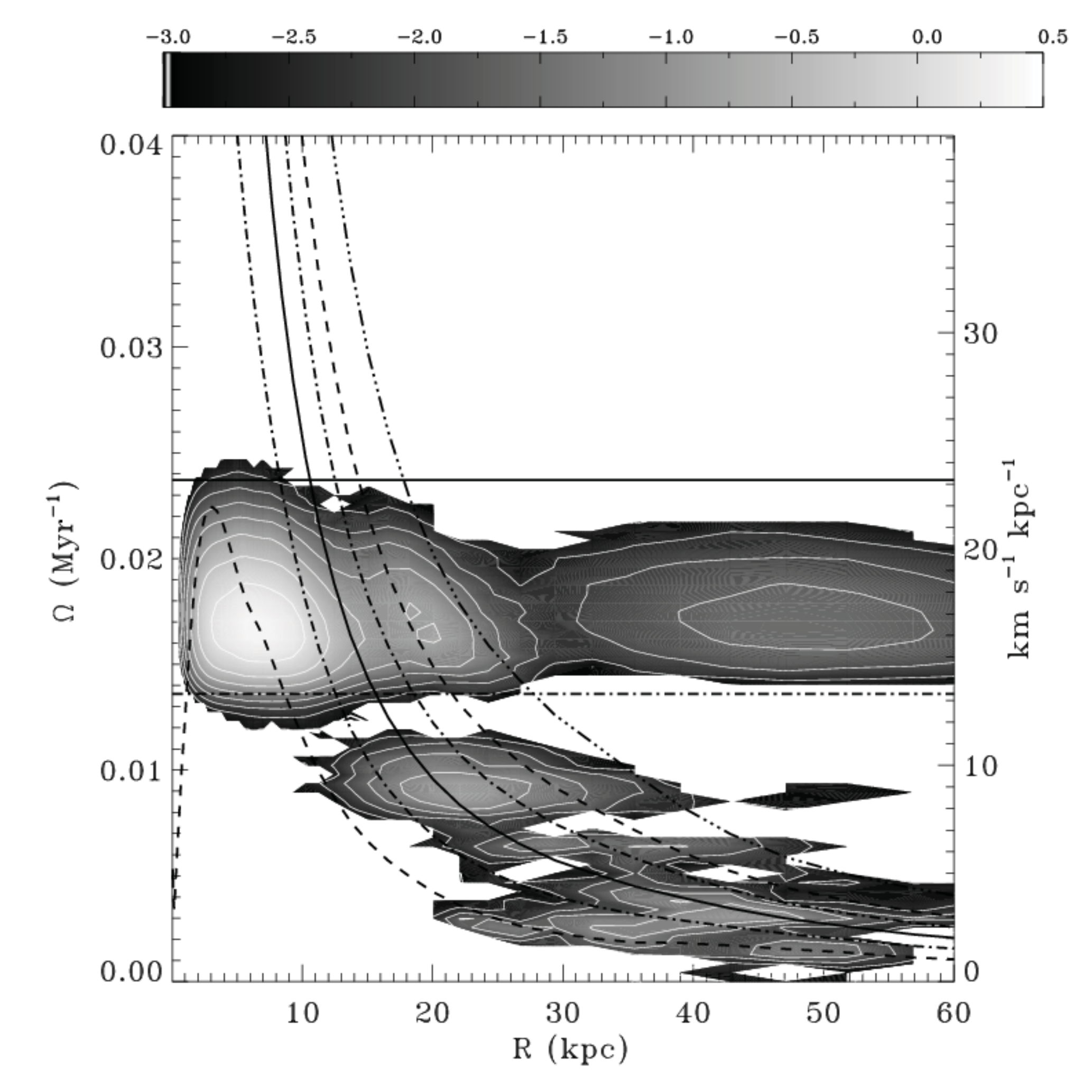}
  \caption{$m=2$ power spectra in $\log$ scale as a function of radius
    for \BCPA\ (left), \FBFH\ (middle) and \PFAA\ (right) in $\log$
    scale. The time interval is $2-10$~Gyr for \BCPA\ and \FBFH\ and
    $3-10$~Gyr for \PFAA. The vertical scales give values of $\Omega$
    in Myr$^{-1}$ (left) and in \omegaunits\ (right).  The averaged
    curves $\Omega -\kappa/2$ (which allows the ILR to be identified)
    and $\Omega +\kappa/2$ (for the OLR) are drawn as black short
    dashed lines, $\Omega -\kappa/4$ and $\Omega +\kappa/4$ (for
    resp. the UHR and 4/1) as dot-dashed line, and $\Omega$ as a solid
    line (for the CR). The horizontal lines represent \Omegap,
    respectively at the beginning of the time window (full line) and
    the end (dot-dashed line). }
  \label{fig:omega_init}
\end{figure*}

By looking for potential differences between these simulations, we can
focus on power spectra of the $m=2$ frequencies between $t=2.11$ and
$10.54$~Gyr for \BCPA\ and \FBFH\ and between $t=3.16$ and
$t=10.54$~Gyr for \PFAA\ (cf Sect.~\ref{sec:chirikov_results} for
explanation of these time ranges).  In Fig.~\ref{fig:omega_init}, the
bar frequency largely dominates.  Several other patterns exist and
give rise to overlaps of resonances. These overlaps are usually
temporary because the resonance system linked to the bar slides
outwards during the evolution of the disc and the slowing down of the
bar. Moreover, since the time window is wide, only long-lived
structures appear in this figure. Transient structures, often with a
lifetime of less than one orbital period, are erased. However, their
role is essential.  In a future article, I will analyze more closely
their connections with more permanent structures.

\section{Chirikov diffusion rate in N$-$body experiments}
\label{sec:chirikov_results}
\begin{figure*}[htb!]
  \centering
  \includegraphics[keepaspectratio,width=0.33\hsize]{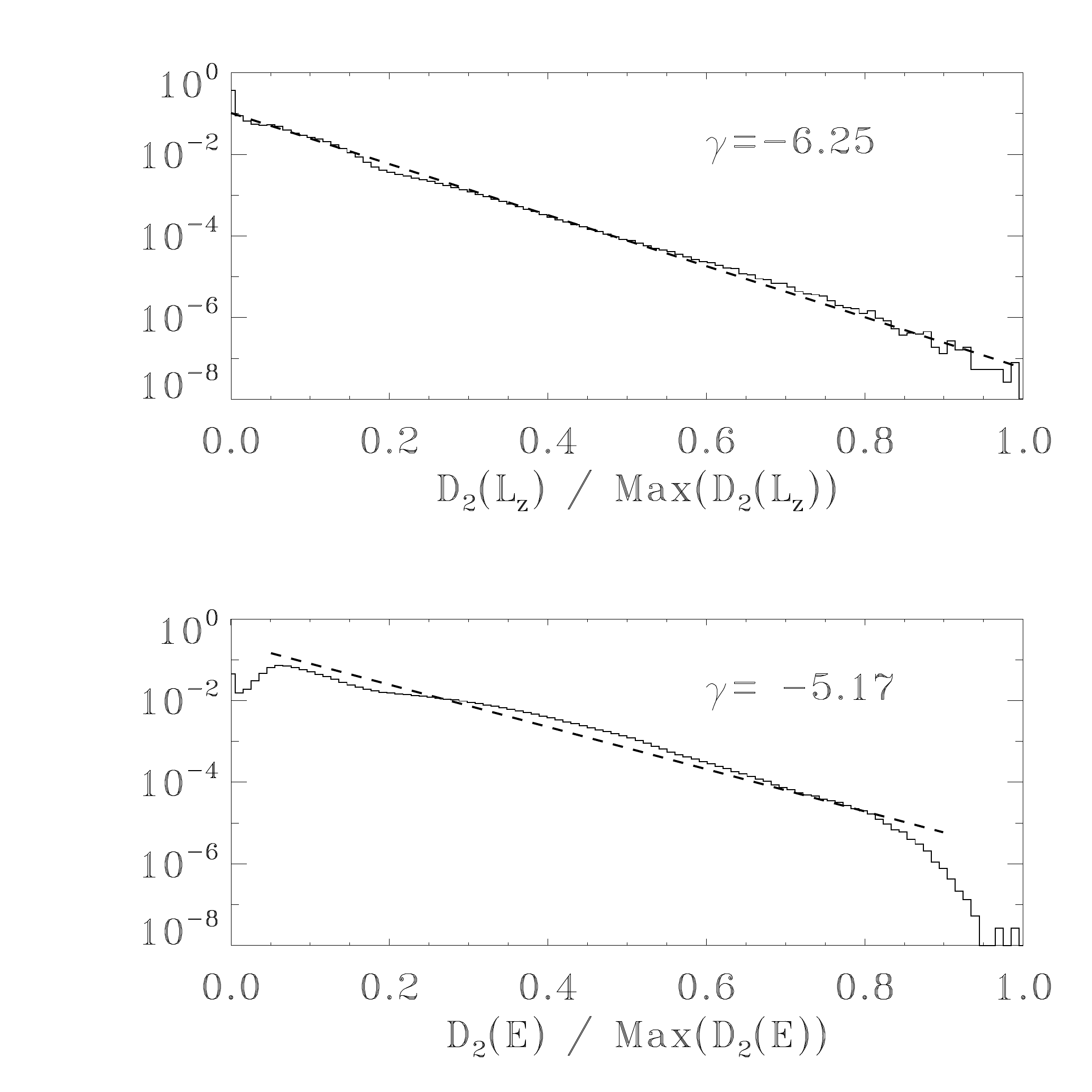}
  \includegraphics[keepaspectratio,width=0.33\hsize]{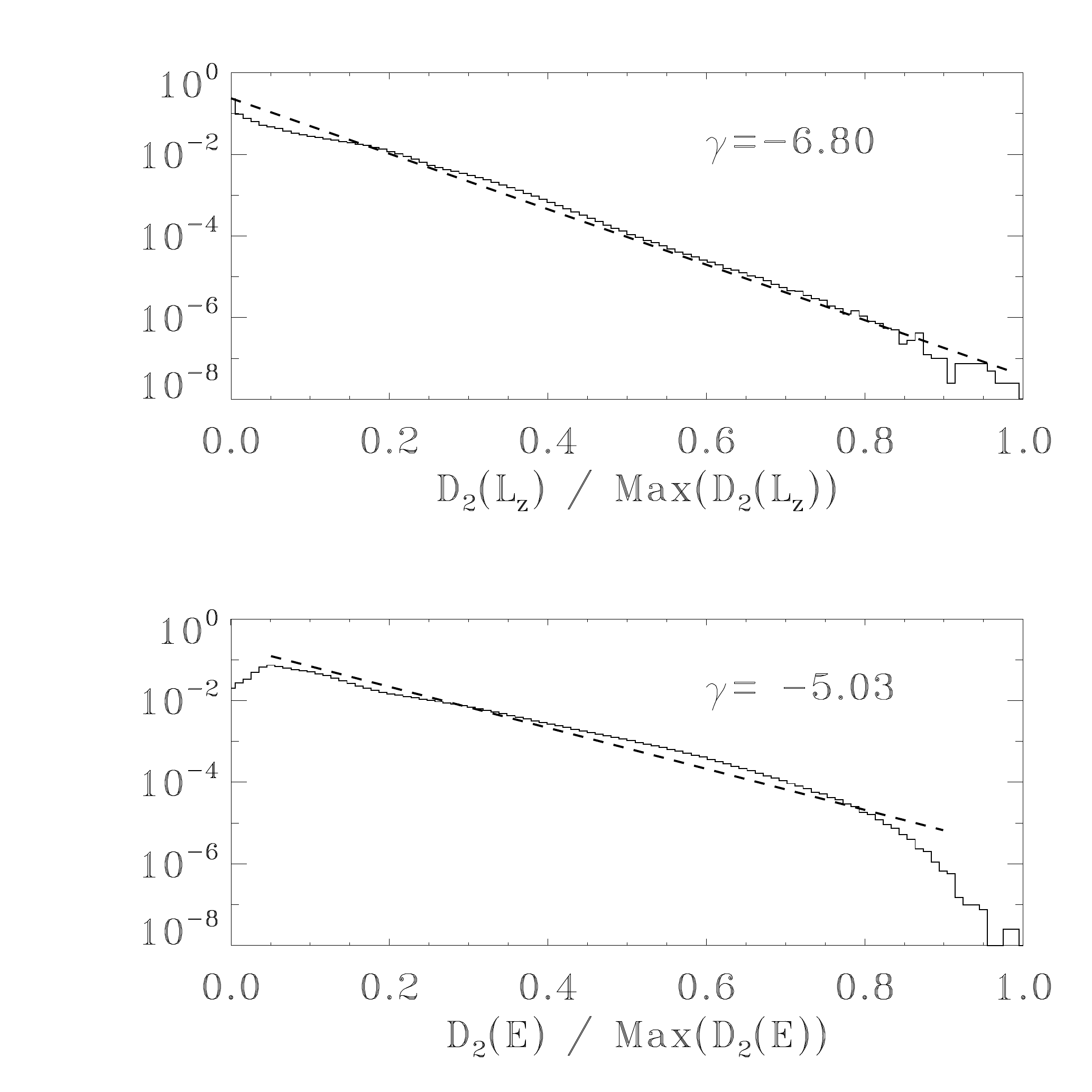}
  \includegraphics[keepaspectratio,width=0.33\hsize]{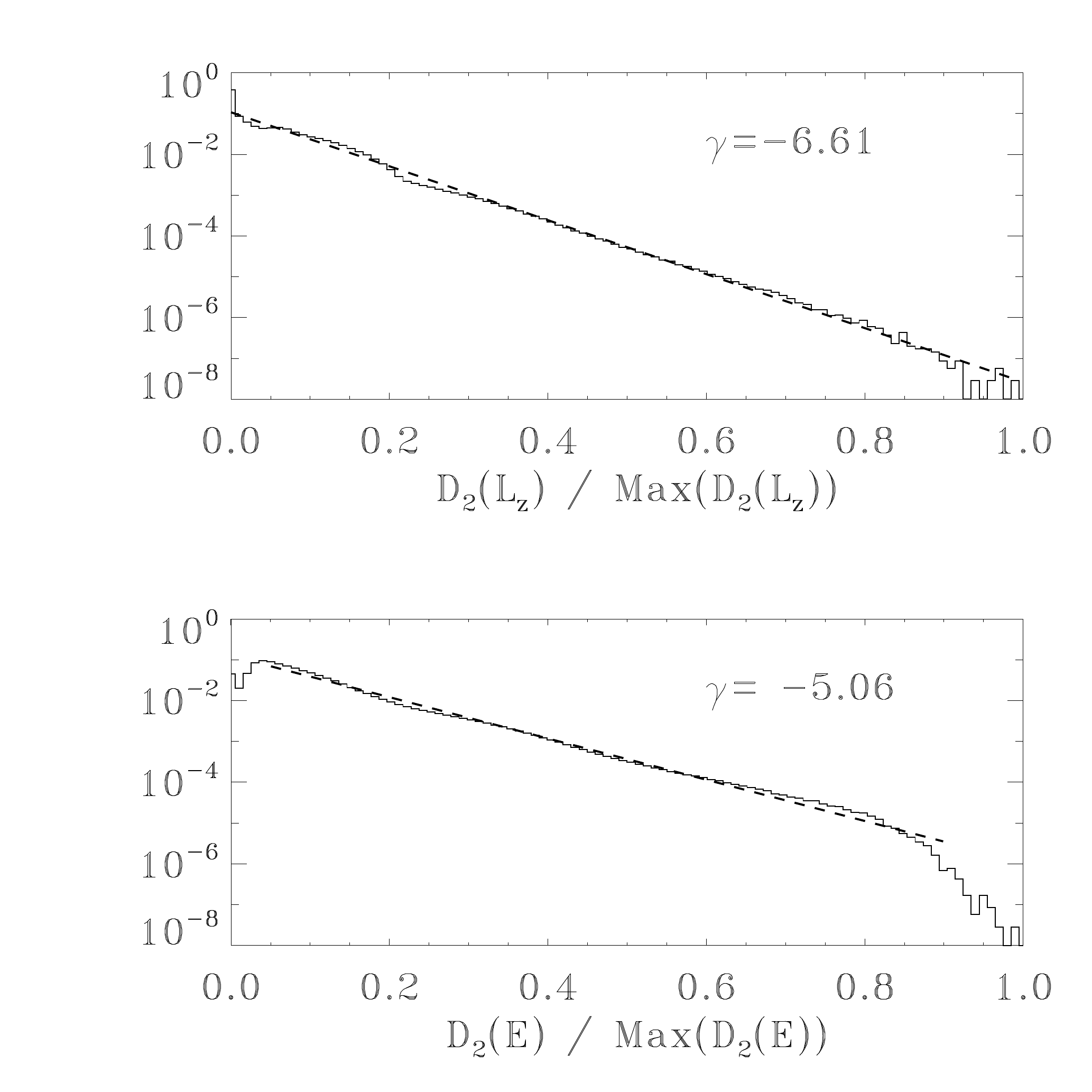}
  \caption{Particle number (or mass fraction) as a function of \dem\ and \dlzm\ for
    \BCPA\ (left), \FBFH\ (middle), and \PFAA\ (right). For
    convenience, the number of particles is normalised to
    $N_\mathrm{s}$ and \dem\ and \dlzm\ to their respective
    maxima. $\gamma$ is the slope fitted by linear regression
    represented by dashed lines. The extent of the dashed line
    represents the range over which the fit has been made.}
    \label{fig:chirikov}
\end{figure*}

As \dem\ and \dlzm\ sum up all fluctuations occurring in the disc, we
have intentionally restricted the time interval to the epoch well
after the bar formation. Therefore, the computation of $D_2(E)$ was
performed between $t=3.16$ and $t=10.54$~Gyr for \PFAA\ (i.e.70
snapshots) and between $t=2.11$ and $10.54$~Gyr for \BCPA\ and
\FBFH\ (i.e. 80 snapshots), spaced by $\Delta t_2$, i.e. on resp. 2415
and 3160 unique pairs. The starting times have been chosen in order to
avoid the strong perturbations caused by the formation of the bar,
which are not of interest to us here.  This rules out the strong
redistribution in $E$ and \Lz\ made by the formation of the bar. Doing
so, we can examine the impact of driving forces in a quieter phase of
the galaxy.

Although particles escaping the grid are tracked throughout their
trajectory by a ballistic approximation, we chose to exclude them from
our analyses as soon as they came out of even one time step. This
drastic procedure ensures that we limit numerical errors to their
lowest values.

Figure~\ref{fig:chirikov} shows the distributions for \dem\ and
\dlzm\ for the three simulations. A first lesson that can be drawn
from these figures is the universality of the distributions shape when
$D_2$ is normalised to its maximum. Approximatively,
$$
\log n(D_2)\propto \gamma D_2 / \max(D_2)
$$ where $\gamma$ is different for \dem\ and \dlzm, and $n(D_2)$ is
the fraction of particles number or, equivalently, the mass
fraction. $\gamma$ is close to $-5.0$ for \dem\ and between $-6.28$
and $-6.79$ for \dlzm.  The shape is represented by a linear
regression valid over a wider range of $D_2/\max(D_2)$ for \Lz\ than
for $E$.

Deviations from a linear fit are also instructive. For \dem, two
regions deserve to be commented on. The three distributions show a dip
for $D_2(E) / \max(D_2(E)) < 0.04-0.06$. It accounts for a maximum of
30\%\ of the total mass. The second region is at the opposite: the
distribution drops when \dem\ reach $\approx 80$~\%\ of the maximum.

However, the normalisation by $\max(D_2)$, which allows to compare the
profiles between them, masks an important element. Indeed, these
maxima are different from one simulation to another, in a sensitive
way because they approximately scale with the square of the total
energy or angular momentum. Table~\ref{tab:max} gives the values of
these maxima. 

\begin{table}[htb!]
\caption{Values of the maxima of \dem\ and \dlzm, and total mass.}
\label{tab:max}
\flushleft
\begin{tabular}{@{}lrll@{}}
\tableline 
Model &  $\max(D_2(E))$ & $\max(D_2(L_\mathrm{z}))$ & $M_\mathrm{tot}$\cr
      &  \demunit\      & \dlzmunit\              & \msol\         \cr
\tableline
\BCPA & $3.93\times 10^{-8}$ & $ 276.7$ & $4.2\times10^{10}$ \cr 
\FBFH & $1.43\times 10^{-7}$ & $ 859.2$ & $1.2\times10^{11}$ \cr 
\PFAA & $3.32\times 10^{-7}$ & $2292.8$ & $2.0\times10^{11}$ \cr 
\PFAB\tablenotemark{a} & $4.46\times 10^{-7}$ & $ 145.6$ & $2.0\times10^{11}$ \cr 
\tableline
\end{tabular}
\tablenotetext{a}{see Section~\ref{sec:axi}}
\end{table}

\section{Diffusion time scales}
\label{sec:timescales}

\begin{figure*}[htb!]
  \centering
  \includegraphics[keepaspectratio,width=0.33\hsize]{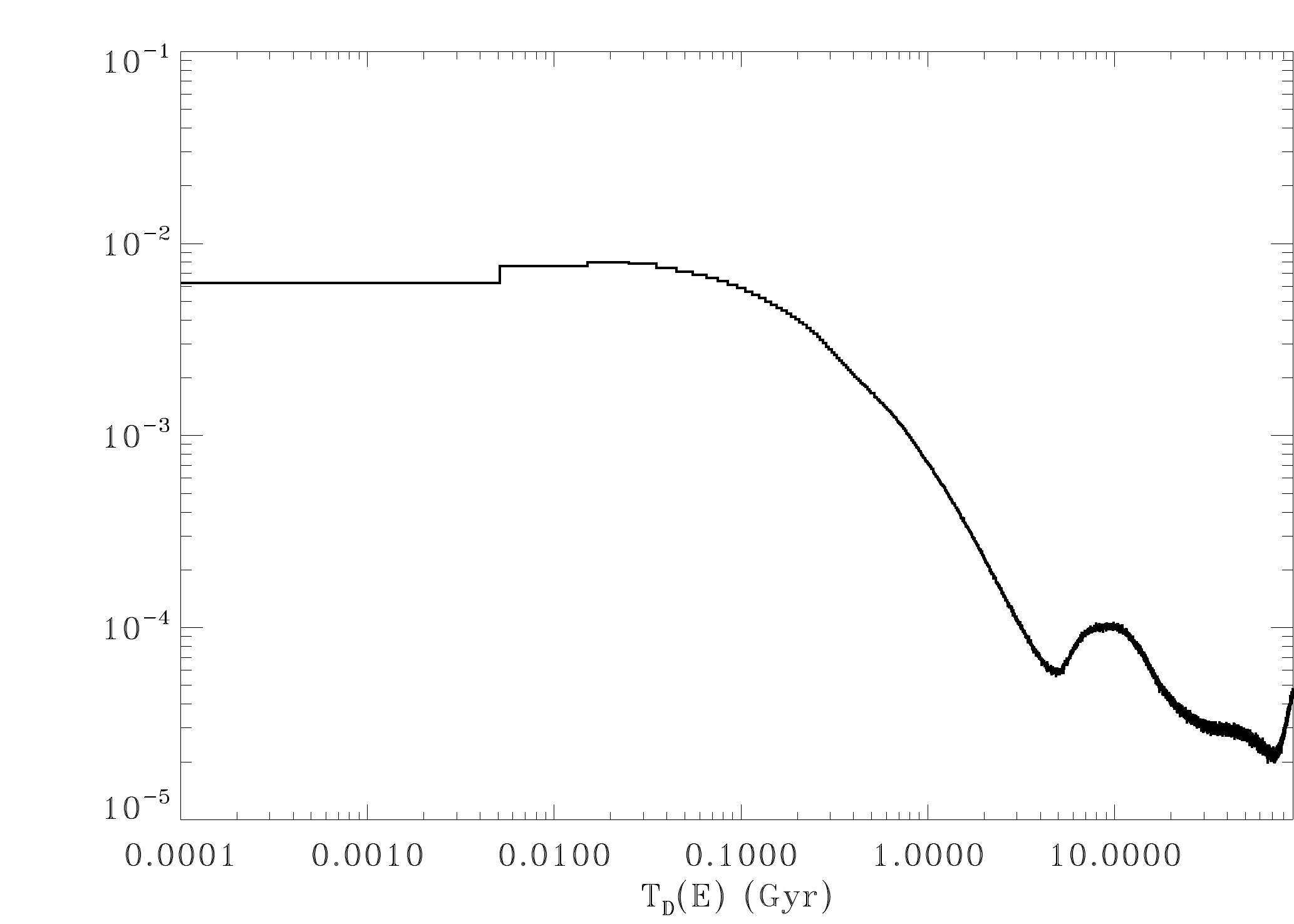}
  \includegraphics[keepaspectratio,width=0.33\hsize]{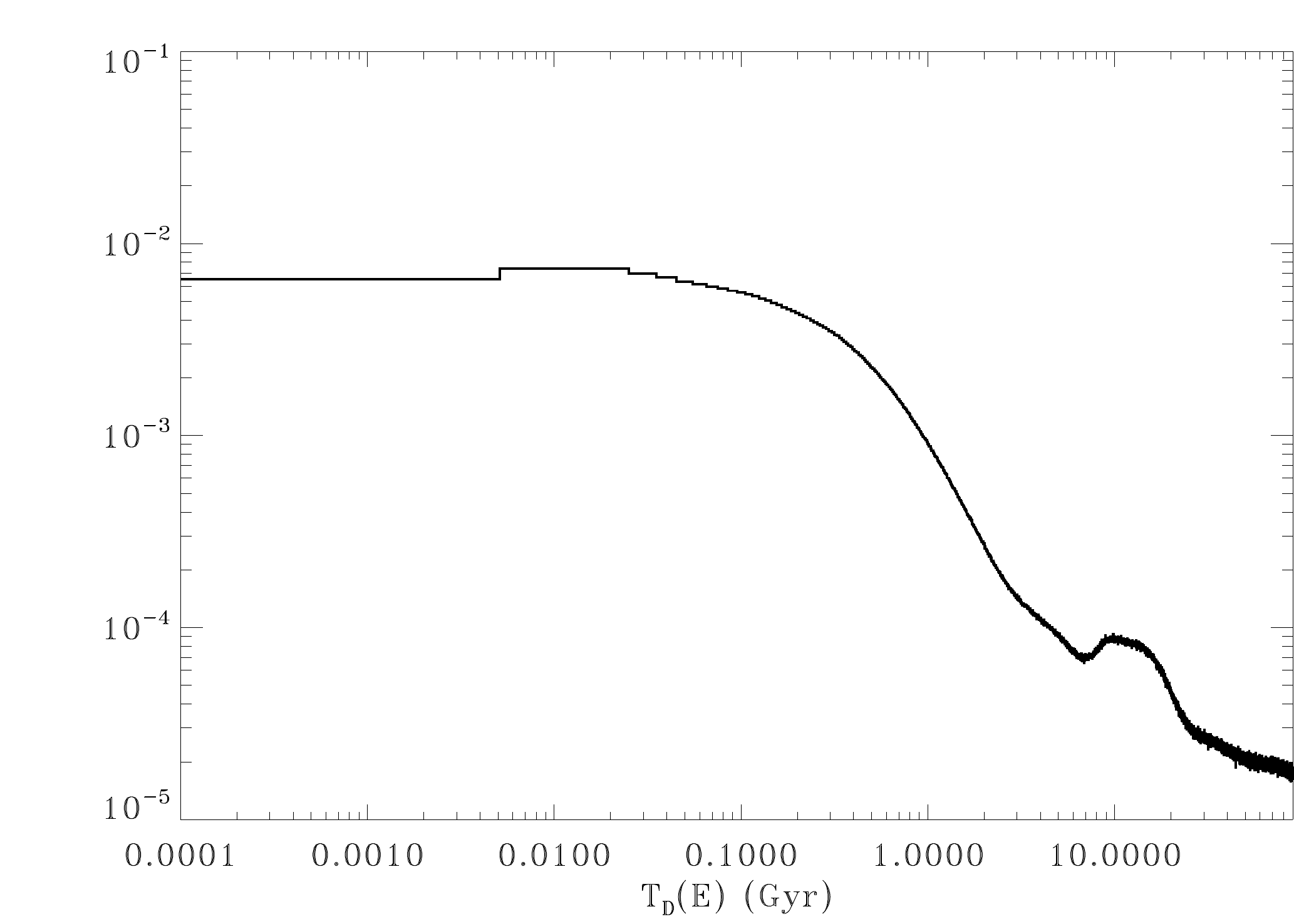}
  \includegraphics[keepaspectratio,width=0.33\hsize]{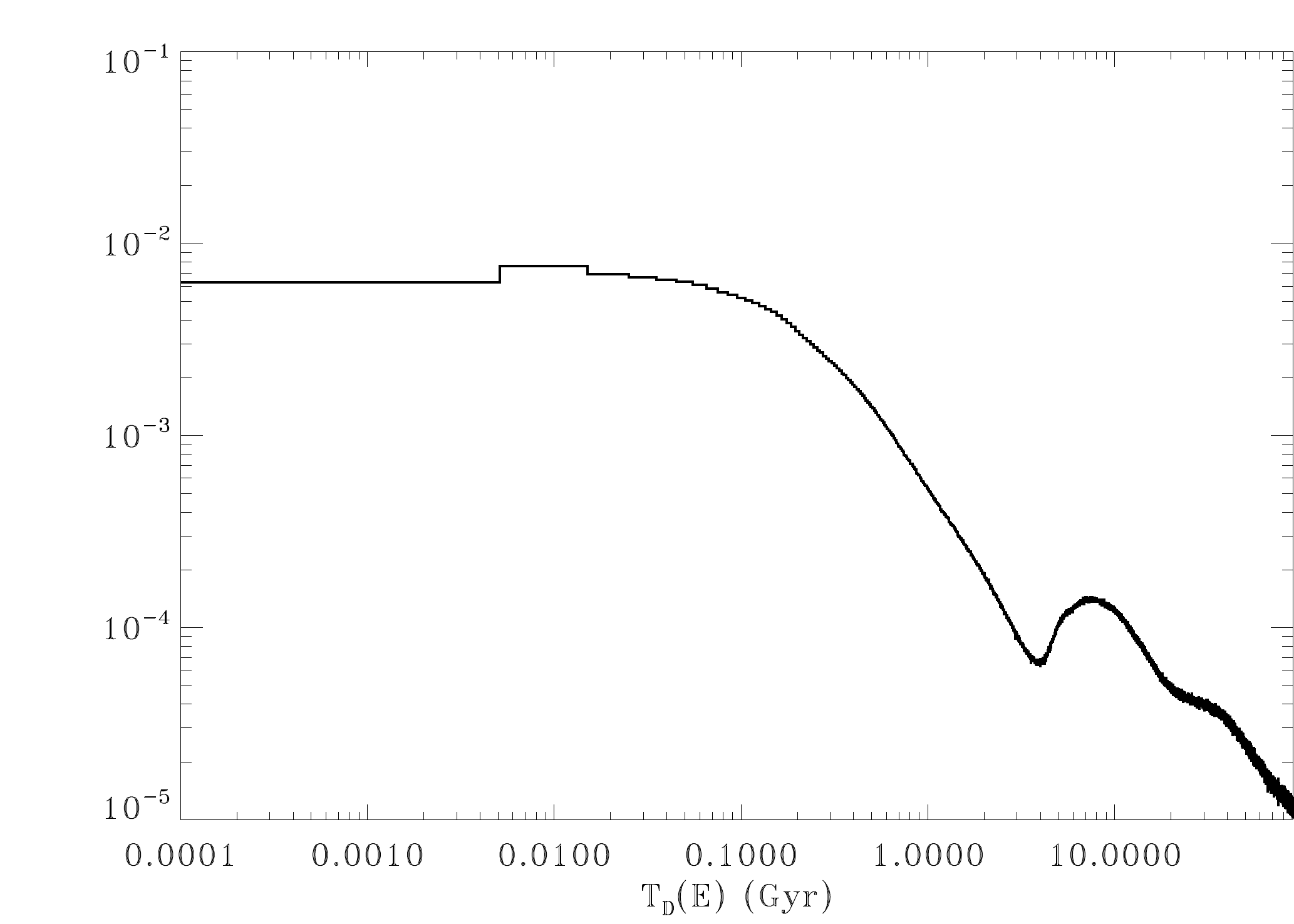}
  \caption{Distribution of particle frequency (or mass fraction) as a
    function of $T_D(E)$ for \BCPA\ (left), \FBFH\ (middle) and
    \PFAA\ (right). 1 particle represents a fraction of $2.5\times
    10^{-8}$ for \BCPA\ and \PFAA, and $2.27\times 10^{-8}$ for
    \FBFH. A binsize of 0.01~Gyr has been used.}
    \label{fig:timescales}
\end{figure*}
\begin{figure*}[htb!]
  \centering
  \includegraphics[keepaspectratio,width=0.33\hsize]{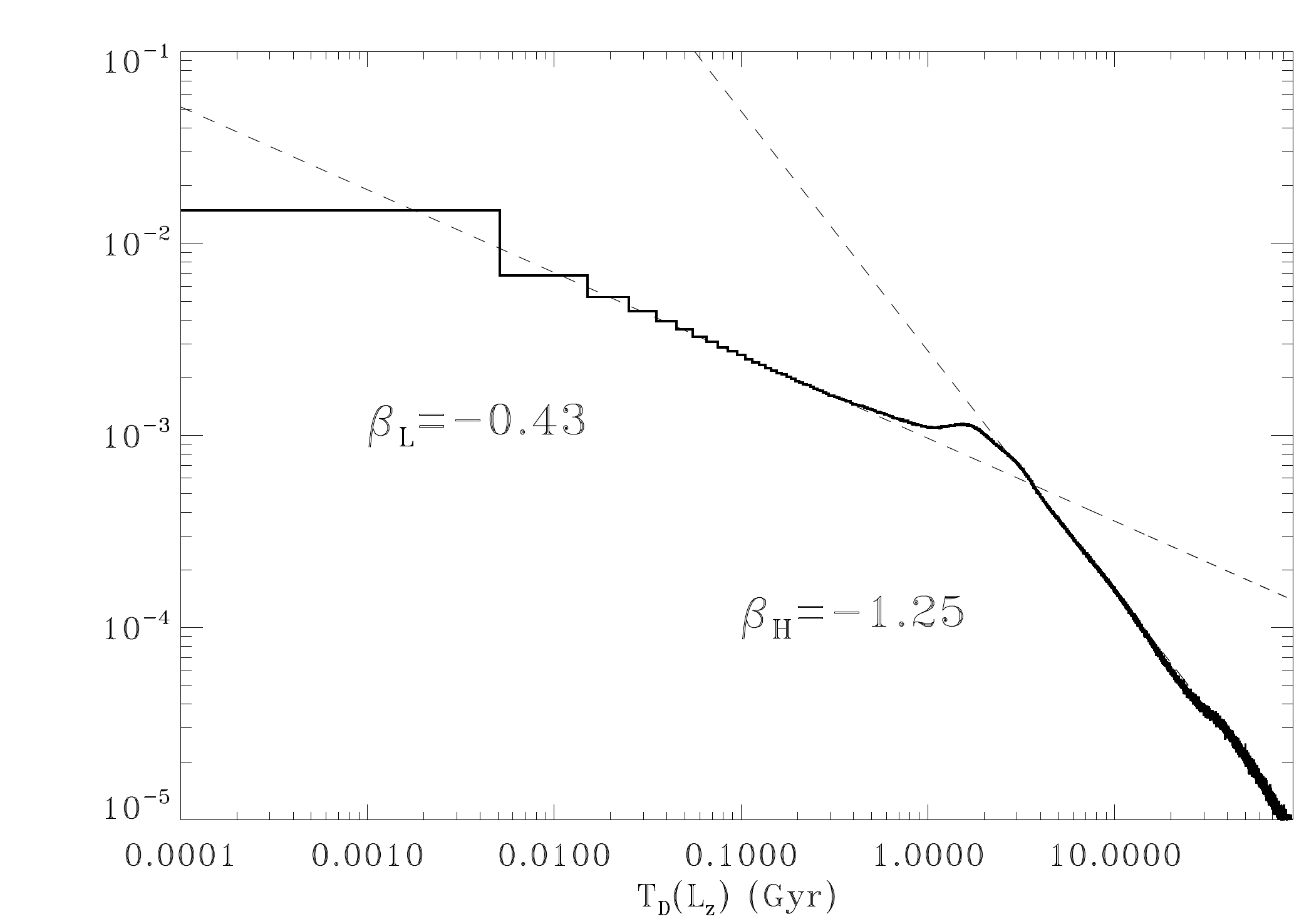}
  \includegraphics[keepaspectratio,width=0.33\hsize]{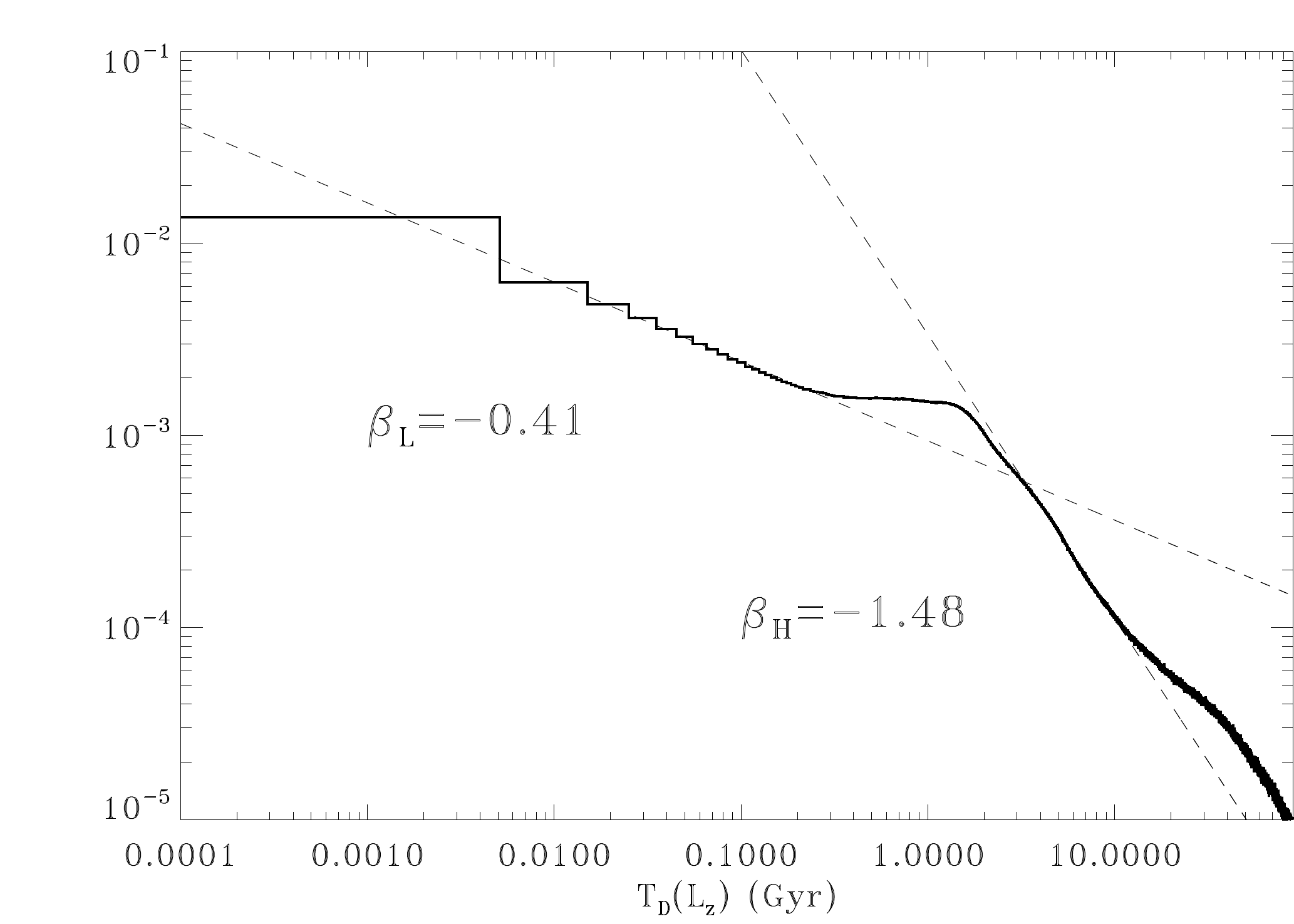}
  \includegraphics[keepaspectratio,width=0.33\hsize]{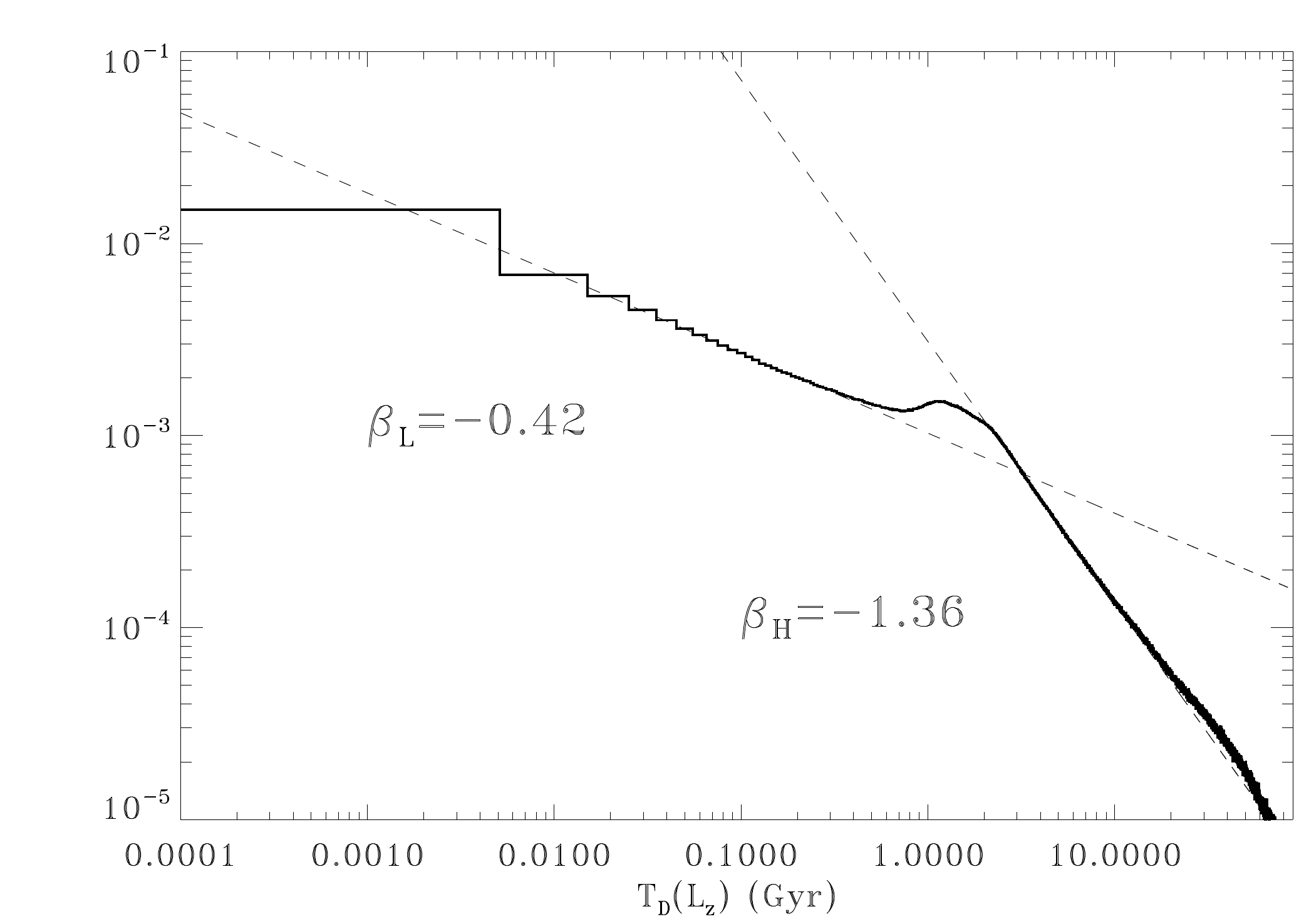}
  \caption{Same as Figure~\ref{fig:timescales} for
    $T_D(L_\mathrm{z})$.}
    \label{fig:timescalesLz}
\end{figure*}

The interpretation of \dem\ may seem complicated because this quantity
mixes information on the quadratic evolution of the $E$ fluctuations
at different time scales. Large fluctuations of $E$ over long times
can contribute as much as small fluctuations over very short times.
Formally, we can also estimate a diffusion time scale $T_D$ by
renormalizing $D_n$ by $E^2$ and $L_\mathrm{z}^2$ respectively. As
Chirikov diffusion rate takes care of all sources of perturbation,
such as particle-wave interactions, it can be seen as a generalisation
of several diffusion time definitions, such as
\citet[]{1942psd..book.....C}'s one on the two-body relaxation times
of stellar systems.

The diffusion time scale, defined as: 
\begin{equation}
T_D(E) = \overline{E}^2 / D_2(E)
\end{equation}
for each individual body, may thus seem more intuitive. The same
definition holds with \Lz\ to compute
$T_D(L_\mathrm{z})$. $E^2(t\!=\!0)$ or $L^2_\mathrm{z}(t\!=\!0)$ can
be used instead of respectively $\overline{E}^2$ or
$\overline{L_\mathrm{z}}^2$ without any significant change. The
results for all three simulations are displayed in
Figure~\ref{fig:timescales} and Figure~\ref{fig:timescalesLz}, where
the frequency distribution of particles (or mass fraction since all
particles have the same individual mass) is plotted against $T_D(E)$
and $T_D(L_\mathrm{z})$. For the sake of clarity, we have restricted
these figures to the range $10^{-4} - 90$~Gyr, but $T_D$ can reach
much higher values for a few particles.

An obvious outcome is the similarity between the distributions for the
three simulations. This form of universality is primarily linked to
the similarities of \dem\ and \dlzm\ distributions for the three
simulations. It is also due to the shape of the distribution functions
DF($E$) and DF(\Lz) (Figure~\ref{fig:fdlzinit}) which, although they
differ in detail, share the same form. Moreover, it should be stressed
here that the time scale chosen is absolute, in Gyr, and not
normalised to a maximum as we have done in
Figure~\ref{fig:chirikov}. Time scales are thus quantitatively comparable in
terms of values.

\subsection{$E$ diffusion time scale}

Dealing first with $T_D(E)$ distribution only, a first region appears
between 0.1~Myr (the minimum time step observed during numerical
integration) and a local minimum located at $\approx 5$~Gyr (\BCPA),
$\approx 7$~Gyr (\FBFH), and $\approx 4$~Gyr (\PFAA).  This time range
covers most of the dynamical time scales present in the galaxy's disc,
from its central part to its outermost border. The decrease in mass
fraction as a function of $T_D(E)$ is slower than exponential.  This
region represents roughly 33\%\ (\BCPA), 39\%\ (\FBFH) and
27\%\ (\PFAA) of the total mass. It is noteworthy that particles with
$T_D(E) \lesssim 0.1$~Myr represent a negligible mass, but $\approx
35-42$\%\ of the total mass lie in the range $T_D(E) < 10.54$~Gyr.
Apart from the fact that $T_D(E)$ is calculated on 1 Gyr more for
\BCPA\ and \FBFH\ than for \PFAA, we did not find any other simple
possible cause that would explain these differences in mass
fraction. For instance, we do not see any scaling with the total mass
or the initial disc scale length. Differences in the evolution of
these three simulations, notably the formation of the bar, the
emergence of the spiral arms, etc., are possibly at the origin of
these differences in mass fraction.

A second feature, a bump centred at $\approx 10$~Gyr for \BCPA\ and
\FBFH, and $\approx 8$~Gyr for \PFAA, might be the footprint of the
limited time length of the simulations. On the contrary, no signature
due to sampling is detected (i.e. 100 Myr for the calculation of
$D_2$).

Finally, $\approx 58-65$\%\ of particles have $T_D(E) >
10.54$~Gyr. This means that most of the mass undergoes energy
fluctuations that only become significant over times longer than the
simulations length, and therefore, in practice, over times that might
be greater than the age of the Universe.

\begin{figure}[htb!]
  \centering
  \includegraphics[keepaspectratio,width=0.75\hsize]{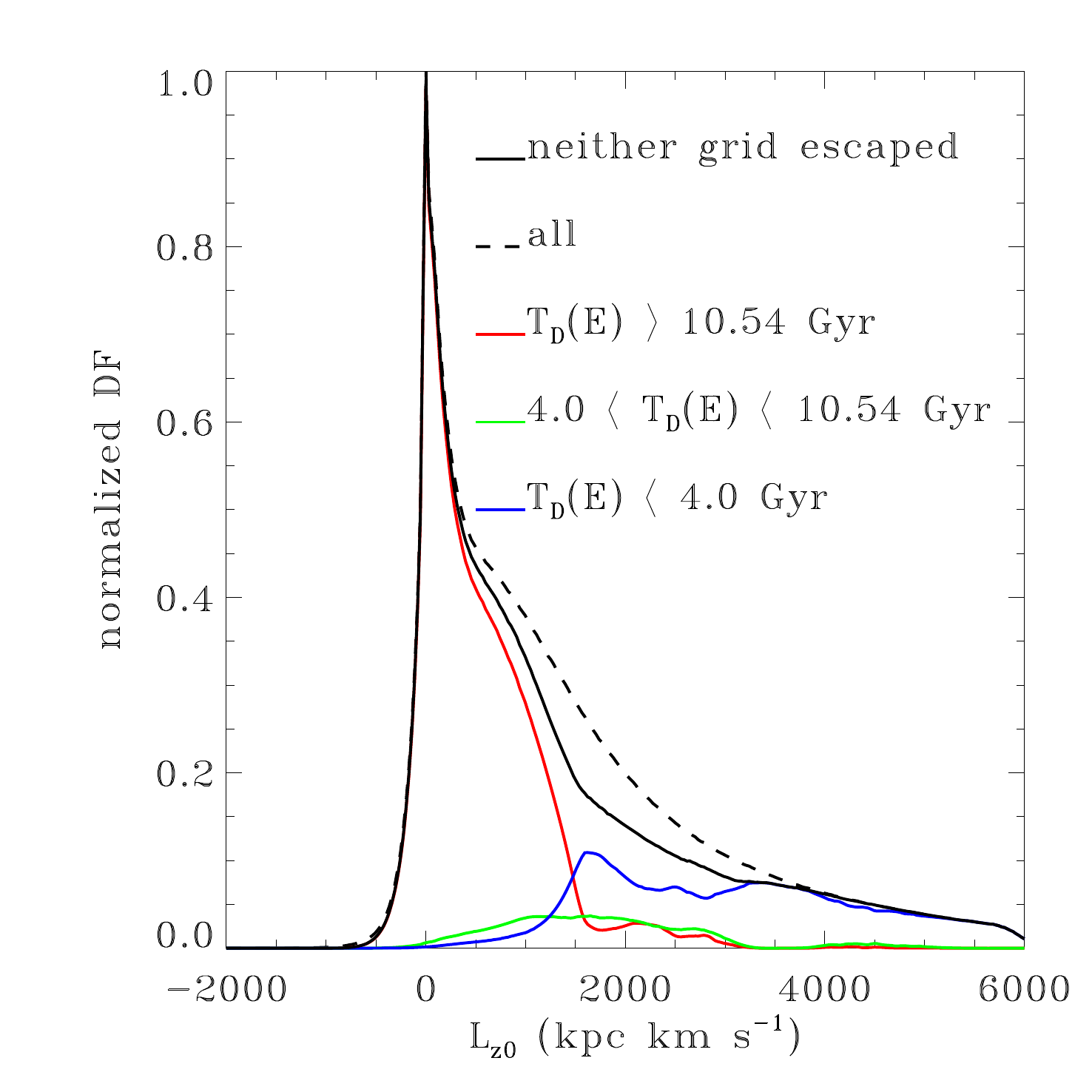}
  \caption{Distribution function DF(\Lz) for \PFAA\ at $t=0$. Dashed
    line: all particles ($N_\mathrm{s}$). Solid line: only particles
    that neither escaped from the grid and used for $T_D(E)$
    computations. Red line: particles with $T_D(E) > 10.54$~Gyr. Green
    line: $4 < T_D(E) < 10.54$~Gyr. Blue line: $T_D(E) < 4.0$~Gyr. All
    DFs have been normalised to DF(0) with all particles, which is
    also the maximum.}
    \label{fig:tde_fdlz}
\end{figure}

In order to understand the properties of the particle populations that
contribute to the different time scales, we have plotted in
Figure~\ref{fig:tde_fdlz} DF(\Lz) for various selection of particles
made on $T_D(E)$ for \PFAA\ (cf. appendix for other simulations). The
reference time is the origin of the simulation ($t=0$). Particles with
$T_D(E) > 10.54$~Gyr come essentially from populations with
$L_\mathrm{z} < 1500$~\Lzunit (red curve in Figure~\ref{fig:tde_fdlz})
that are typically well inside the innermost resonances of the
bar. For the sake of comparison, an hypothetical circular orbit at the
bar Ultra-Harmonic resonance (UHR) at $t=3.16$~Gyr has $L_\mathrm{z}
\approx 2200$~\Lzunit.  Although the bar is a major gravitational
perturbation, which has the ability to cause significant mass
redistribution, the fact that resonances isolate the central region
from the rest of the galaxy possibly limits the onset of energy
diffusion. Therefore, the diffusion time scales in $E$ are longer than
simulation length in the innermost region.

Particles with $4 < T_D(E) < 10.54$~Gyr (green curve in
Figure~\ref{fig:tde_fdlz}) come from a fraction of the bar population
which is close to the corotation barrier. This region contains many
bifurcations of orbit families by period doubling
\citep{1983ApJ...275..511C}. An infinite cascade of this type of
bifurcation then forms a sequence that leads to stochasticity.

Finally, particles with $T_D(E) < 4$~Gyr come massively
from both the `hot' population and the disc. Their diffusion time
scale is comparable to or shorter than typical dynamical time scales
in the disc. 

We can therefore summarise the global trend of $T_D(E)$ by saying that
it decreases from the centre to the most external regions. This trend
will be further discussed in Sect.~\ref{sec:discussion} and refined.

\subsection{\Lz\ diffusion time scale}

Dealing now with $T_D(L_\mathrm{z})$, a noteworthy observation is that
two slopes appear for $0.001 \lesssim T_D(L_\mathrm{z}) \lesssim 0.3$~Gyr, and
$3\lesssim T_D(L_\mathrm{z}) \lesssim 10$~Gyr in a $\log-\log$ diagram.  These
ranges are those on which a line has been fitted by a standard
algorithm of linear regression. Therefore, for $0.001 \lesssim
T_D(L_\mathrm{z}) \lesssim 0.3$~Gyr,
\begin{equation}
\label{eq:ntd}
n(T_D) \propto T_D^{\beta_\mathrm{L}}
\end{equation} 
where $\beta_\mathrm{L} \approx -0.42$ in most cases and $n(T_D)$ is
the distribution of particle frequencies (or mass fraction).  The
second slope with index $\beta_\mathrm{H}$ seems to depend on the
simulation parameters.

\begin{figure}[htb!]
  \centering
  \includegraphics[keepaspectratio,width=0.75\hsize]{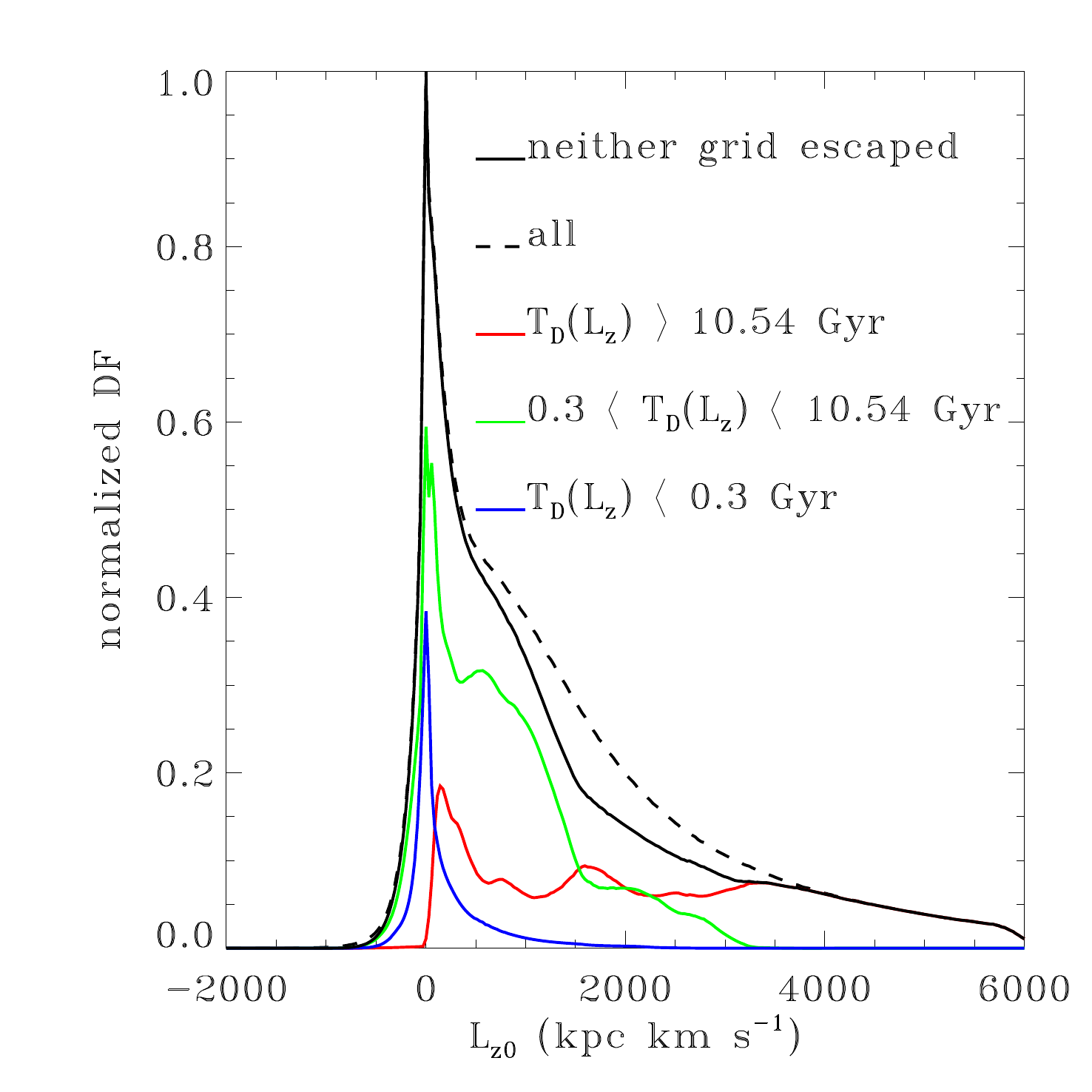}
  \caption{Same as Figure~\ref{fig:tde_fdlz} but for
    $T_D(L_\mathrm{z})$. The noticeable time scale is now 0.3~Gyr
    instead of 4~Gyr.}
    \label{fig:tdlz_fdlz}
\end{figure}

Below the time scale of 0.3~Gyr, which is also the typical mean bar
rotation period for all three simulations, mass is made of particles
with low \Lz\ (Figure~\ref{fig:tdlz_fdlz}
for the case of \PFAA). These particles represent only a small mass
fraction, between 8.6 and 9.5\%.

On the other side of the distribution, 36-40\% of the mass has
$T_D(L_\mathrm{z}) \ge 10.54$~Gyr. All kinds of orbits contribute to
this population, but it should be noted that, in the case of \PFAA,
all particles with $L_\mathrm{z} > 3300$~\Lzunit, i.e. a large
fraction of the `hot' population and all disc particles, have very
long $T_D(L_\mathrm{z})$.

Finally, about half of the mass (51-54\%) has intermediate diffusion
times, between 0.3 and 10.54~Gyr.  The particles inside the bar form
the largest part of this population responsible for the diffusion of
the angular momentum. Probably a small fraction of the `hot'
population also belongs to this category but it is difficult to
quantify its contribution more precisely without a detailed orbit
analysis that is postponed to a future paper.

In comparison, the global trend of $T_D(L_\mathrm{z})$ seems to be
opposite to that of $T_D(E)$: the diffusion time scale increases with
the radius.

\section{The axisymmetric case}
\label{sec:axi}

An instructive element of comparison is to look at what happens to
$D_2(E)$, $D_2(L_\mathrm{z})$, $T_D(E)$, and $T_D(L_\mathrm{z})$ in
case a simulation is forced to remain axisymmetric. Both $E$ and
\Lz\ are now isolating integral of motion. Diffusion rates would be
zero if the gravitational potential were due to an infinite number of
particles. The potential would then be smooth and stationary. The
individual energy of the particles would then be perfectly
preserved. Poissonian shot noise due to potential discreteness, forces
accuracy and the finite number of particles is however unavoidable. We
thus need reference values.

\begin{figure}[htb!]
  \centering
  \includegraphics[keepaspectratio,width=\hsize]{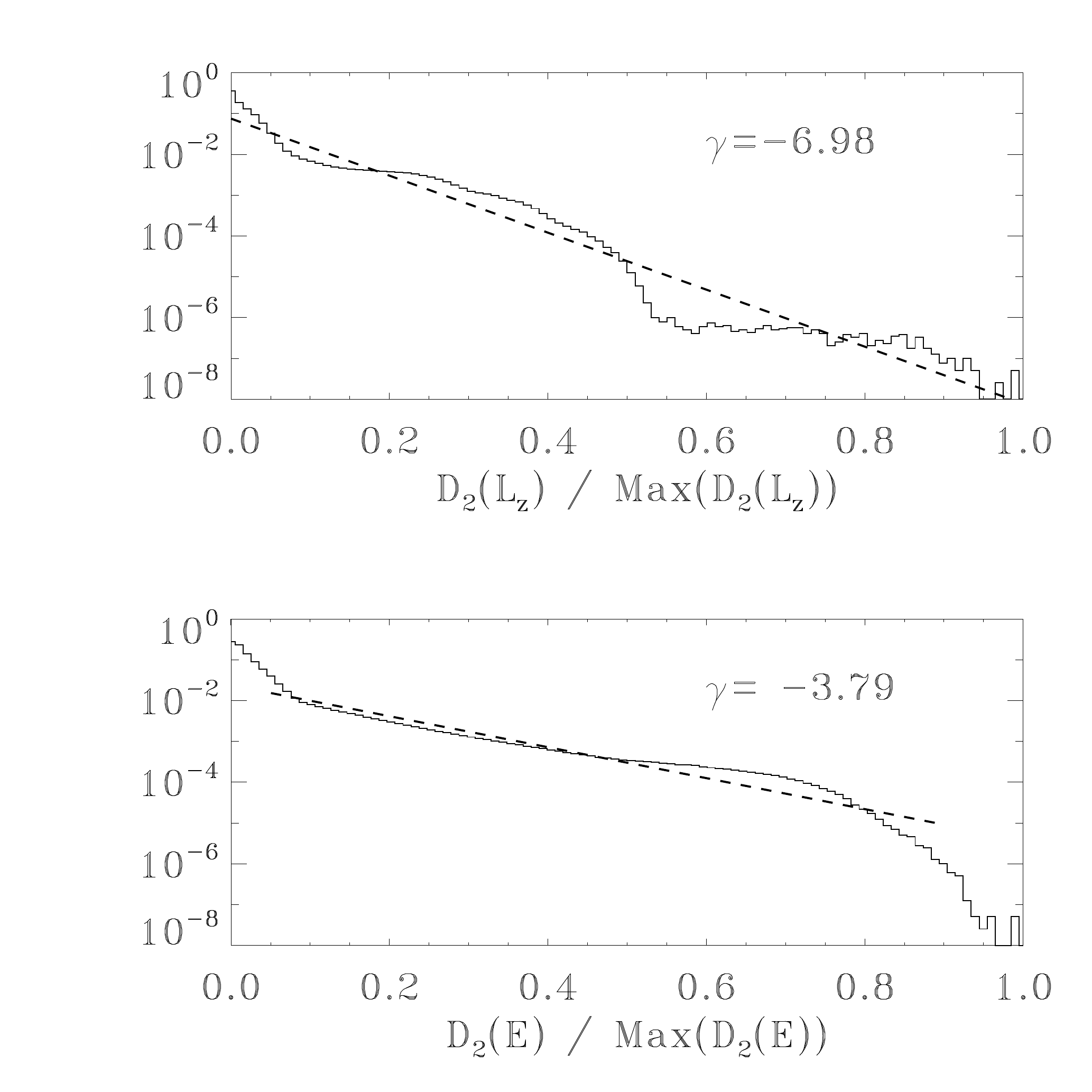}
  \caption{Particle number (or mass fraction) as a function of \dem\ and \dlzm\ for
    \PFAB\ when the mass density is forced to remain axisymmetric. As
    for Figure~\ref{fig:chirikov}, the number of particles is
    normalised to $N_\mathrm{s}$ and \dem\ and \dlzm\ to their
    respective maxima (see text for values).}
    \label{fig:chirikov_axi}
\end{figure}

\PFAA\ was recalculated by forcing the axisymmetrisation of the mass
density at each time step, any other parameter being similar to
\PFAA. Let us call it \PFAB. The gravitational potential therefore
remains close to axisymmetric, no bar or spiral structure can develop.
Only axisymmetric waves can propagate in the first Gyr of the
simulation, carrying initial angular momentum towards the external
regions. For comparison purposes, all rates and time scales were
calculated in the same way as \PFAA, i.e. between $t=3.16$ and
$t=10.54$~Gyr.

Regarding $D_2(E)$ (Figure~\ref{fig:chirikov_axi}), even if its
maximum ($4.46\times10^{-7}$) is close but a little greater than that
of \PFAA\ (cf. Table~\ref{tab:max}), the distribution shape is
significantly different. Indeed, a large mass fraction has low values
of $D_2(E)$ (i.e. less than 10\% of the maximum). Beyond that, the
distribution is flatter than for \PFAA, which results in a higher
$\gamma$ slope ($\approx -3.8$ instead of $\approx -5$). 

For $D_2(L_\mathrm{z})$, not only $\max(D_2(L_\mathrm{z}))\approx 146$
is much lower than for \PFAA\ ($\approx 2300$), and this for the same
total mass, but the shape of the distribution is no longer close to a
linear relation between the mass fraction and $D_2/\max(D_2)$. The
scale of $D_2(L_\mathrm{z})$ has been reduced by a factor of 16. This
can be easily understood because a large angular momentum diffusion is
not expected in an axisymmetric simulation as \Lz\ is an integral of
motion.

\begin{figure}[htb!]
  \centering
  \includegraphics[keepaspectratio,width=0.75\hsize]{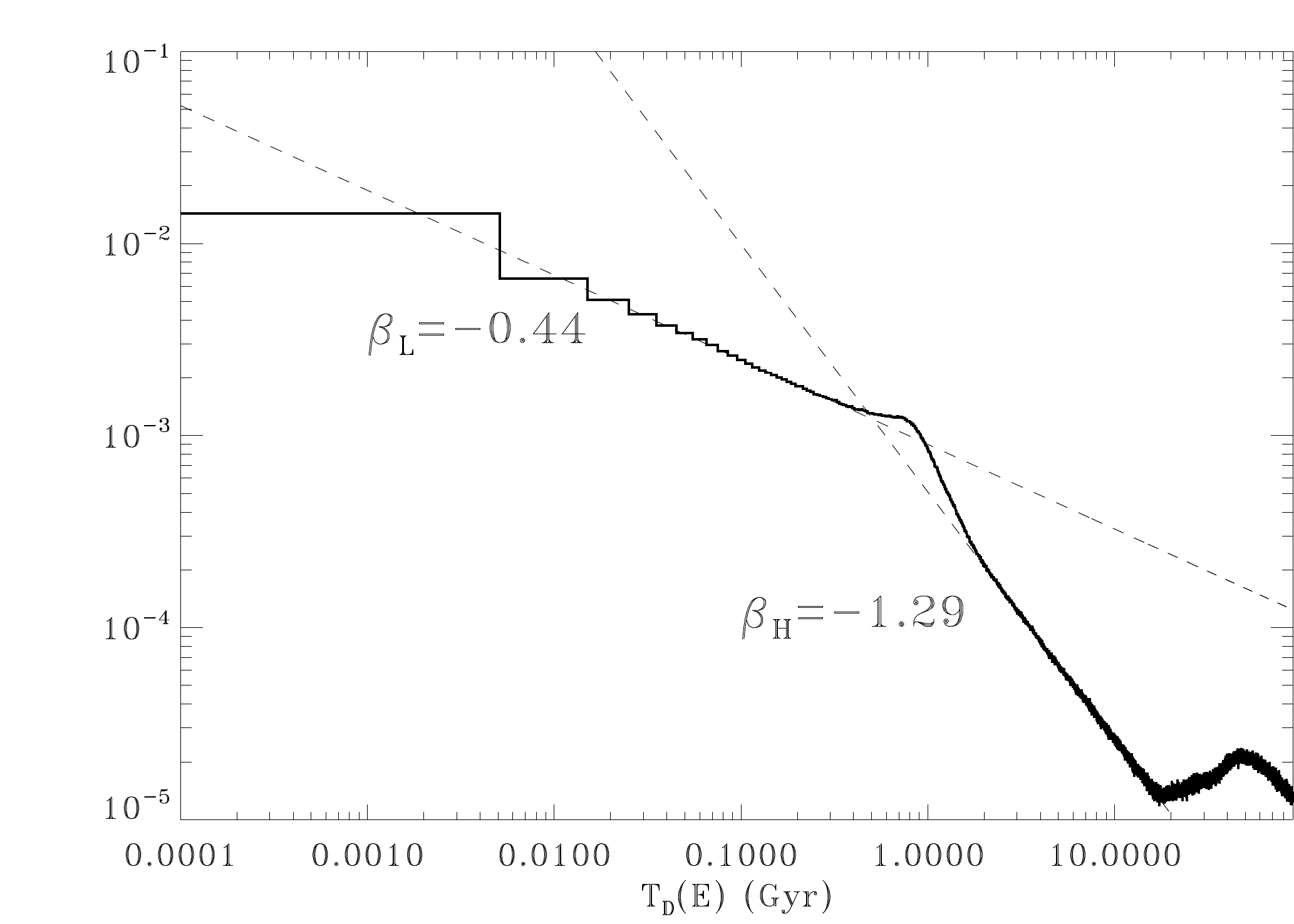}
  \includegraphics[keepaspectratio,width=0.75\hsize]{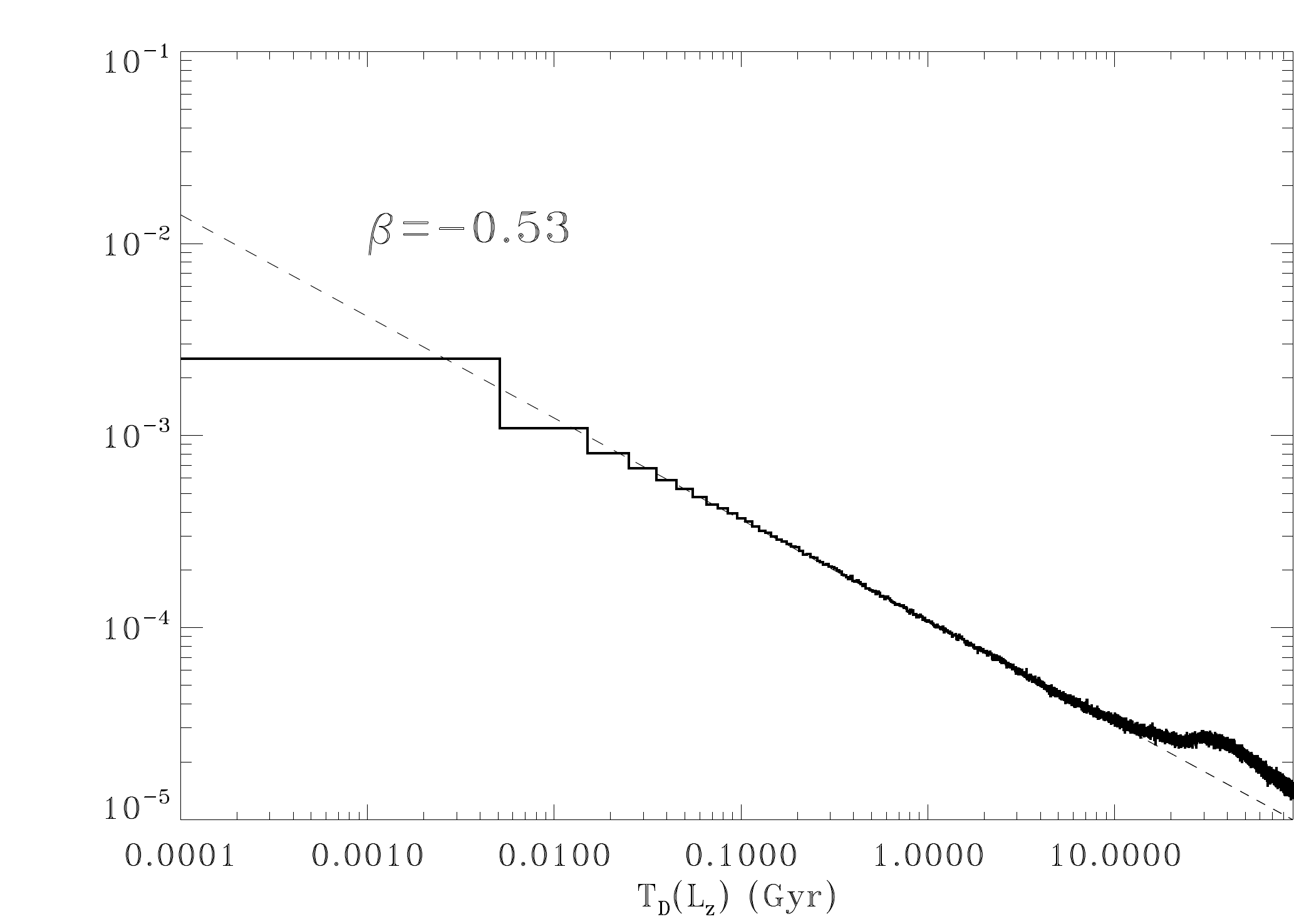}
  \caption{Same as Figure~\ref{fig:timescales} (top) and
    Figure~\ref{fig:timescalesLz} (bottom) for \PFAB\ when the mass
    density is forced to remain axisymmetric. }
    \label{fig:timescales_axi}
\end{figure}

\begin{figure}[htb!]
  \centering
  \includegraphics[keepaspectratio,width=0.75\hsize]{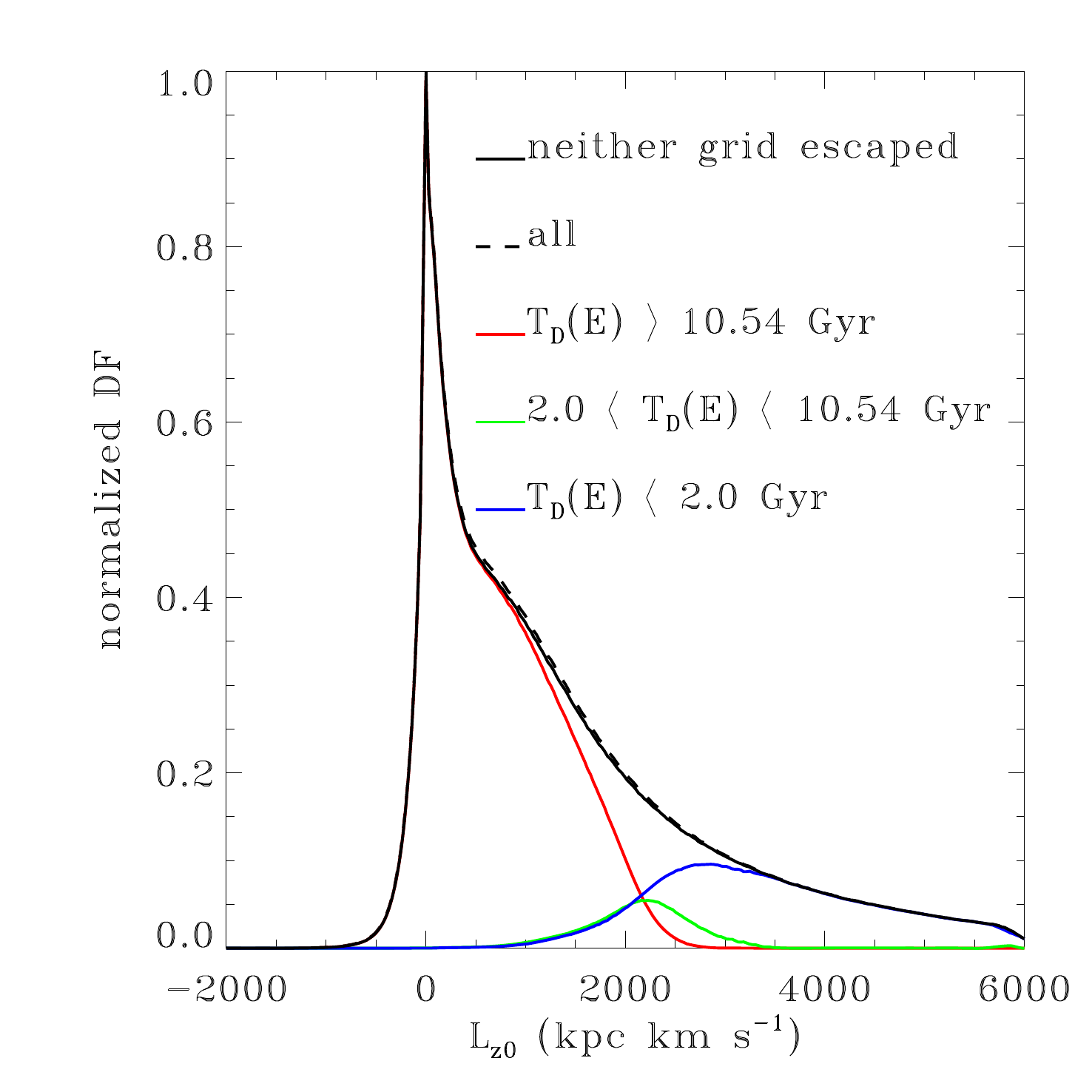}
  \includegraphics[keepaspectratio,width=0.75\hsize]{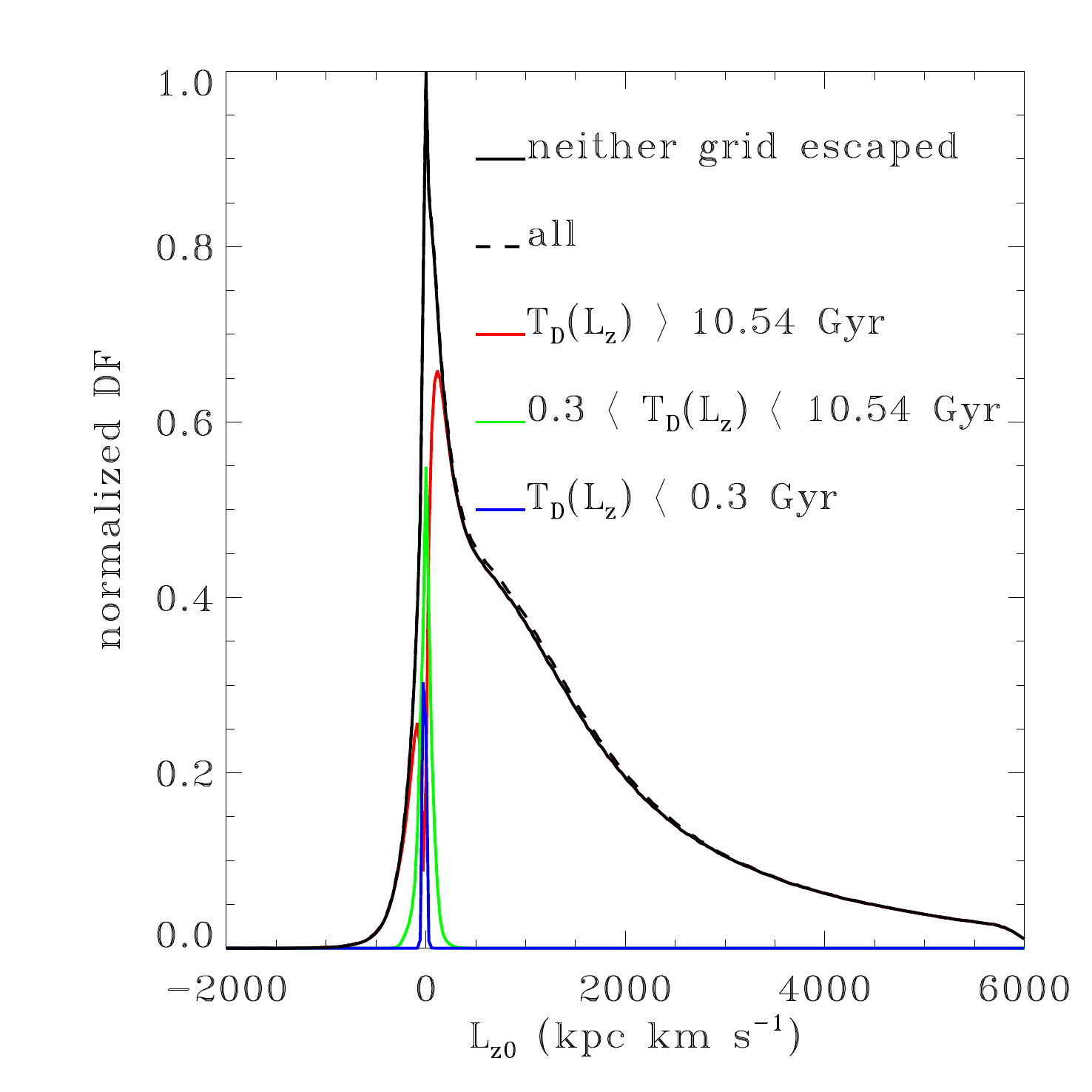}
  \caption{Distribution function DF(\Lz) for
    \PFAB\ at $t=0$. Dashed lines: all particles
    ($N_\mathrm{s}$). Solid lines: only particles that neither escaped
    from the grid and used for $T_D(E)$ and $T_D(L_\mathrm{z})$ computations. Red lines:
    particles with $T_D(E)$ or $T_D(L_\mathrm{z}) > 10.54$~Gyr. Green
    lines: $2 < T_D(E) < 10.54$ (top) or $0.3 < T_D(L_\mathrm{z}) <
    10.54$~Gyr (bottom). Blue lines: $T_D(E) < 2.0$~Gyr (top) or
    $T_D(L_\mathrm{z}) < 0.3$~Gyr (bottom). All DFs have been normalised to
    DF(0) with all particles, which is also the maximum.}
    \label{fig:td_fdlz_axi}
\end{figure}

Unexpectedly, the shape of the $T_D(E)$ distribution
(Figure~\ref{fig:timescales_axi} top row) between 0.1~Myr and 4~Gyr is
significantly different from those shown in
Figure~\ref{fig:timescales}. It looks like the $T_D(L_\mathrm{z})$
distribution for non-axisymmetric simulations. The two slopes
$\beta_\mathrm{L}$ and $\beta_\mathrm{H}$ are moreover similar to
those displayed in Figure~\ref{fig:timescalesLz}. If we select the
particles of \PFAB\ with $T_D(E) < 2$~Gyr, it appears
(Figure~\ref{fig:td_fdlz_axi}) that their \Lz\ corresponds to the
so-called `hot' population, although formally this population cannot
exist here because the bar and associated resonances are absent.
These particles are therefore the ones likely to be most affected by a
perturbation, their diffusion time being already the shortest in the
axisymmetric case. Particles with $ 2 < T_D(E) < 10.54$~Gyr would then
be orbits located in the region inside the corotation, where the
families of orbits undergo bifurcations. Here again, in the
axisymmetric case, and therefore in absence of any pattern frequency,
there is no specific resonances.

The $T_D(L_\mathrm{z})$ distribution of
\PFAB\ (Figure~\ref{fig:timescales_axi} bottom row) displays now a
unique slope with $\beta \approx -0.53$ that extends from 0.01 to
$\approx 10$~Gyr. But particles in this range account for only 7\% of
the total mass compared to more than 50\% for the other three
non-axisymmetric simulations. Only orbits with very low \Lz\ (and thus
close to the centre) contribute to this region. The rest of the mass
(93\%) has $T_D(L_\mathrm{z}) \gg 10$~Gyr. This trend is much expected as
\Lz\ is an integral of motion in axisymmetric discs. 
This illustrate that when a bar, spiral structure, and any other
pattern appear in the \PFAA\ simulation, these collective oscillations
are solely responsible for the diffusion of the angular momentum. As
soon as axisymmetry is broken, $T_D(L_\mathrm{z})$ starts decreasing,
leading to $T_D(L_\mathrm{z}) < 10.54$~Gyr for circular orbits with
$L_\mathrm{z} < L_\mathrm{z}(\mathrm{corotation})$.

A final observation worth mentioning: the number of escaped particles,
defined as those having gone outside the grid even one time step, is
only 1.5\% for \PFAB, compared to 14\% for \PFAA.

\section{Discussion}
\label{sec:discussion}

\subsection{Role of resonances}
At resonances, particles can undergo two types of phenomena that are
both important for the galactic dynamics. The first one is capture (or
trapping) into the resonance. Any particle follows closely an
adiabatic trajectory until it crosses a resonant surface. There, the
particle trajectory starts following this surface rather than the
adiabatic trajectory, causing strong adiabatic invariant (hereinafter
referred to as $I$) changes along the exact motion. As the particle
may escape from the resonance (after a finite but unpredictable time),
it starts following a different adiabatic trajectory with value of $I$
completely different from the initial one. Initial conditions of
particles to be captured and those to cross the resonance without
capture are entangled. At the coarse-grained level, this leads to
capture probabilities.  These capture probabilities are of order of
the perturbation strength so that we do not expect a large part of
particles to be captured. However, due to the disc rotation, phase
trajectories of the averaged system are closed, allowing particles to
cross the resonant surface again and again. This significantly
increases the probability to be captured on the long term, making
resonant capture important for the galactic disc dynamics.

The second phenomenon, scattering, takes place for particles that
cross the resonance without capture. The particle trajectory follows
closely the adiabatic trajectory, but at the resonant surface it
shifts by a very small amount rather than being captured. After
crossing the resonance, the particle follows a new adiabatic
trajectory, which has moved from a distance of order of the
perturbation strength from the original one. The amount by which $I$
changes depends on the initial conditions. As for capture, multiple
scattering is possible and lead to diffusion of the adiabatic
invariant on the long term. Therefore, particles passing through
resonance change their energy even in absence of trapping. However, if
the resonant system contains a separatrix, the mean energy change due
to scattering is finite. Multiple scatterings lead to either
acceleration or deceleration of particles. In absence of such a
separatrix, the energy change over multiple scatterings is diffusive.

In the context of epicyclic approximation, localising resonances
requires the computation of the circular orbit frequency $\Omega$ and
the radial epicyclic frequency $\kappa$.  Strictly speaking, these
frequencies predict the oscillation frequencies of the orbits in the
axisymmetrical case limits only. They do not provide any indication of
whether families of periodic orbits do follow such oscillations when
the bar growth breaks the axisymmetry. However, a number of previous
orbital studies \citep[cf.][and discussion
  therein]{2006A&A...452...97M} suggest that the epicyclic
approximation could lead to an acceptable estimation of the resonance
locations, in particular if we are mainly interested in their
evolution rather than their accurate absolute position. For instance,
using a careful integration of orbits to compute $\Omega$ and
$\kappa$, \citet{2006A&A...452...97M} found that the error on the
corotation radius remains within 10\%.

In the case of barred galaxies, the resonant area is a large region
around the idealised corotation radius, in the sense of a radial
solution of the equation $\Omega(R_\mathrm{cor})=\Omega_p$ as defined
by the linear theory of resonances.  Indeed, Lagrangian points are
defined as being the points of equilibrium between centrifugal and
centripetal forces along the main axis of the bar perturbation. The
radii of these points converge towards the corotation circle when the
bar perturbation vanishes. \rlun\ and \rlquatre\ are the radii of the
Lagrangian points along respectively the major-axis and intermediate
axis (minor-axis in a face-on projection) of the bar perturbation.
\citet[][]{2006A&A...452...97M} have shown that the relative amplitude
of the difference between \rlun\ and \rlquatre\ rarely exceeds
15\%\ even in very strong bar phases. A standard value for a slowly
evolving bar seems to be in the range 5 to 10\%.  Moreover, the
amplitude of the difference between \rlun\ and \rlquatre\ is roughly
proportional to the bar strength. When the bar gets stronger, the
difference between \rlun\ and \rlquatre\ increases, \rlquatre\ being
always smaller than \rlun.  Therefore, the corotation radius always
lies between the Lagrangian points radii but is closer to \rlun\ than
\rlquatre. Thus, when the corotation is mentioned, the region
concerned is an oval ring whose width depends on the strength of the
bar. \citet{2007MNRAS.379.1155C} focused on the action of the bar in
the redistribution of \Lz\ and, in particular, the capture of
particles by corotation. Their work clearly shows that the scope of
the action of the bar goes well beyond corotation. In particular, they
show how strong the variations of \Lz\ of the particles trapped around
stable Lagrange points are.

It is moreover well established that a rotating stellar bar transports
angular momentum outwards, resulting in a decrease in \Omegap. This
decrease is almost linear with time in absence of a dissipative
component (gas) and any star formation. For instance, this is the case
of the three N$-$body simulations described in
Sect.~\ref{sec:nbody}. As a result, the corotation radius increases
over time, as do other resonance radii. It is the whole resonance
system that moves, whether it is the one generated by the bar or by
any other structure likely to lose/gain angular momentum.

\begin{figure*}[htb!]
  \centering
  \includegraphics[keepaspectratio,width=0.24\hsize]{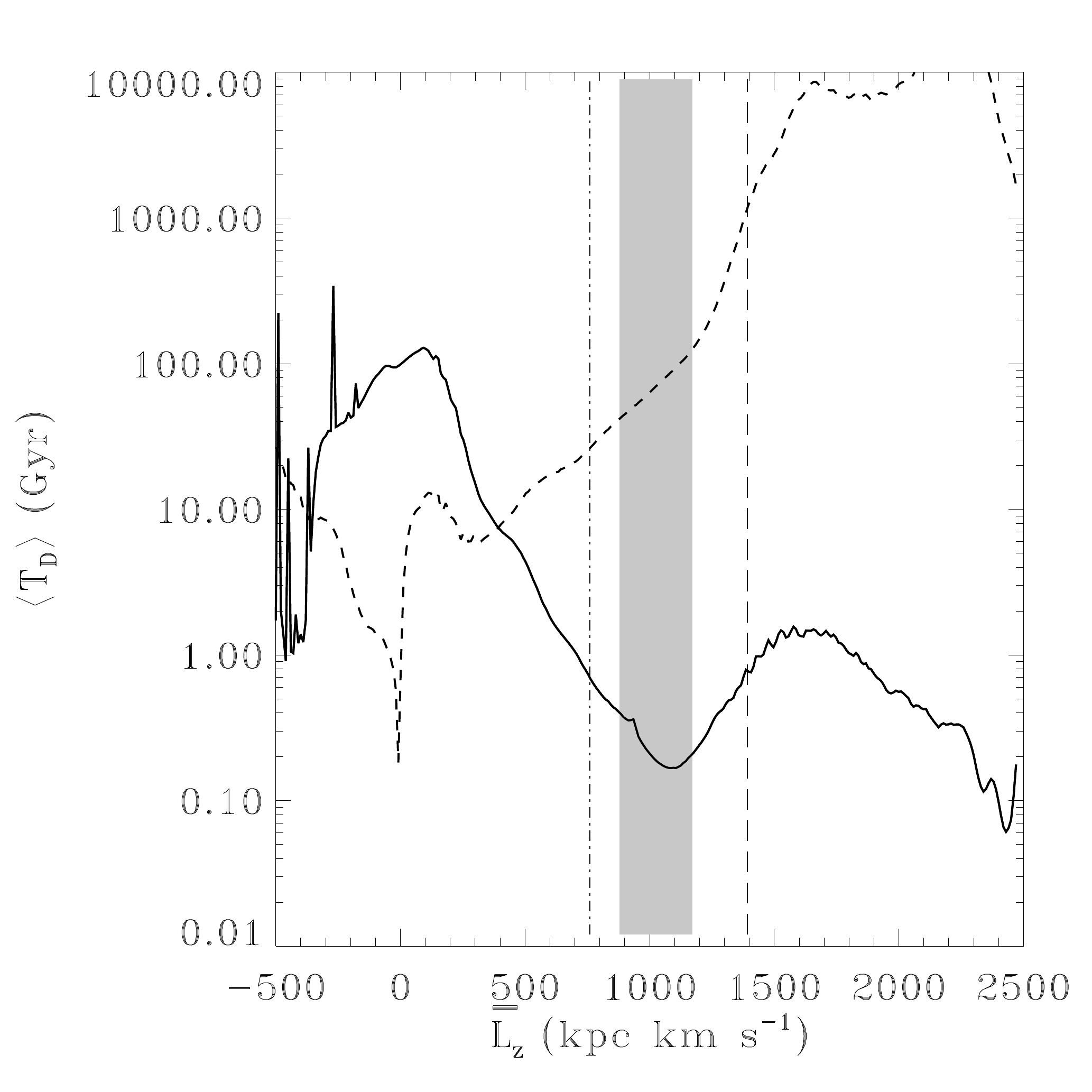}
  \includegraphics[keepaspectratio,width=0.24\hsize]{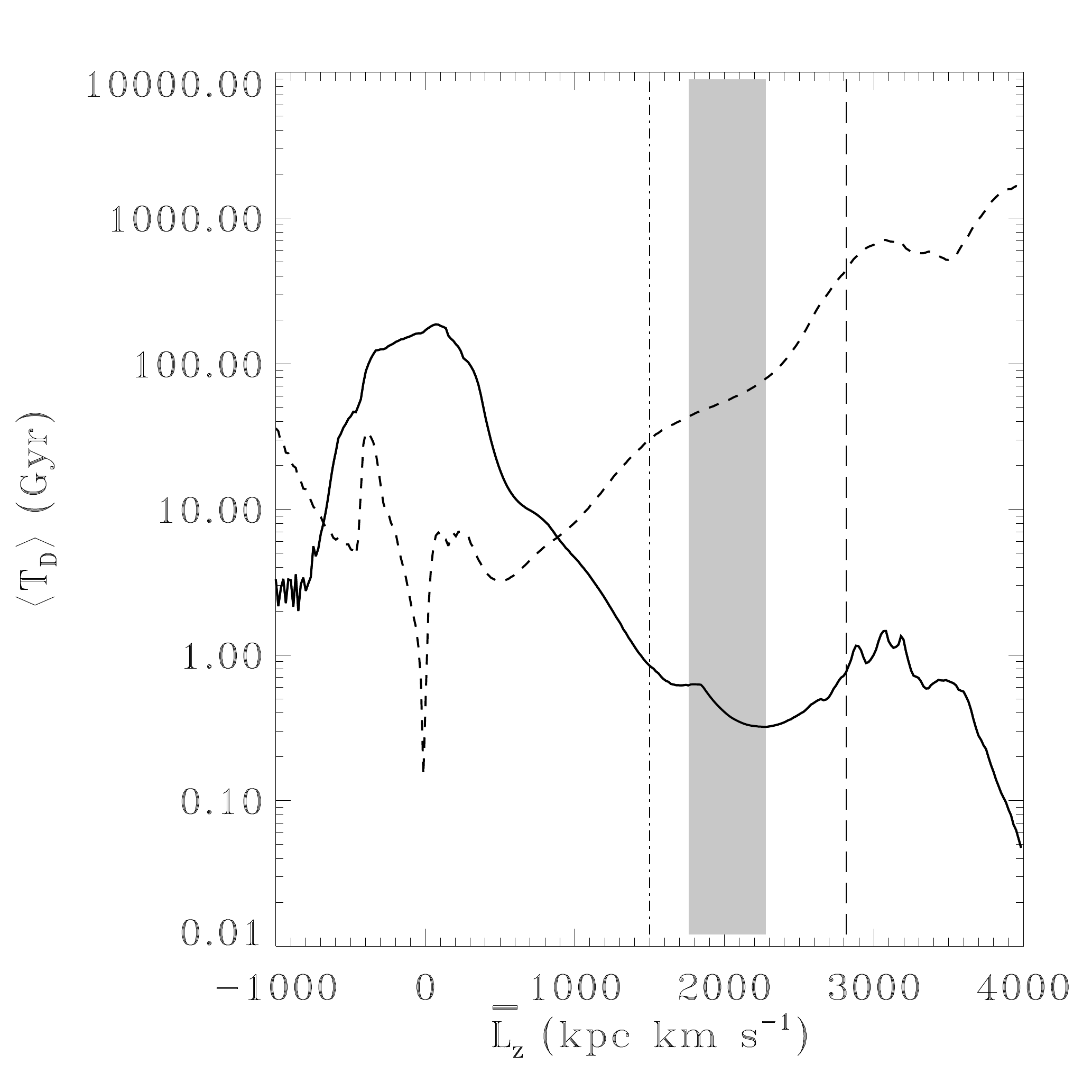}
  \includegraphics[keepaspectratio,width=0.24\hsize]{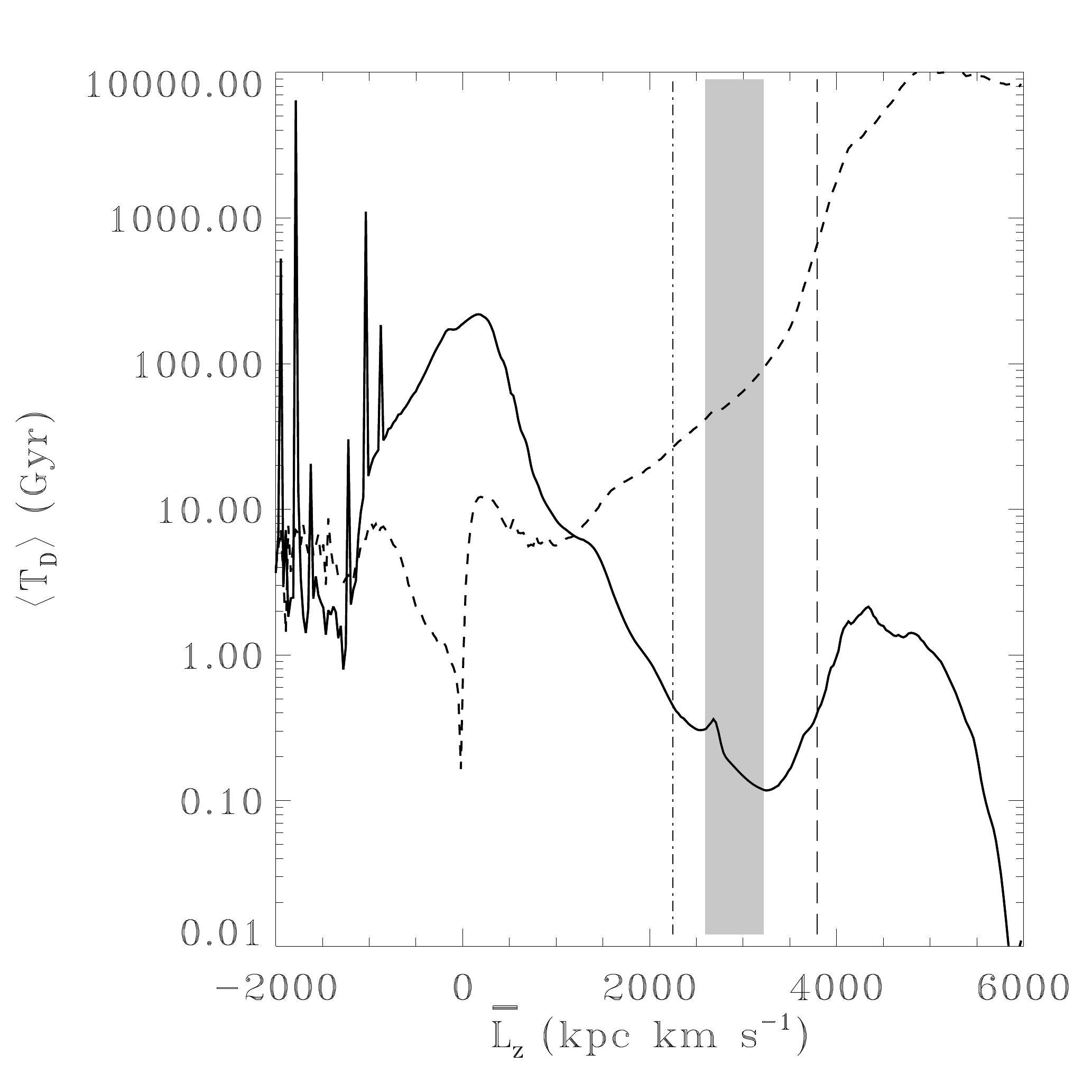}
  \includegraphics[keepaspectratio,width=0.24\hsize]{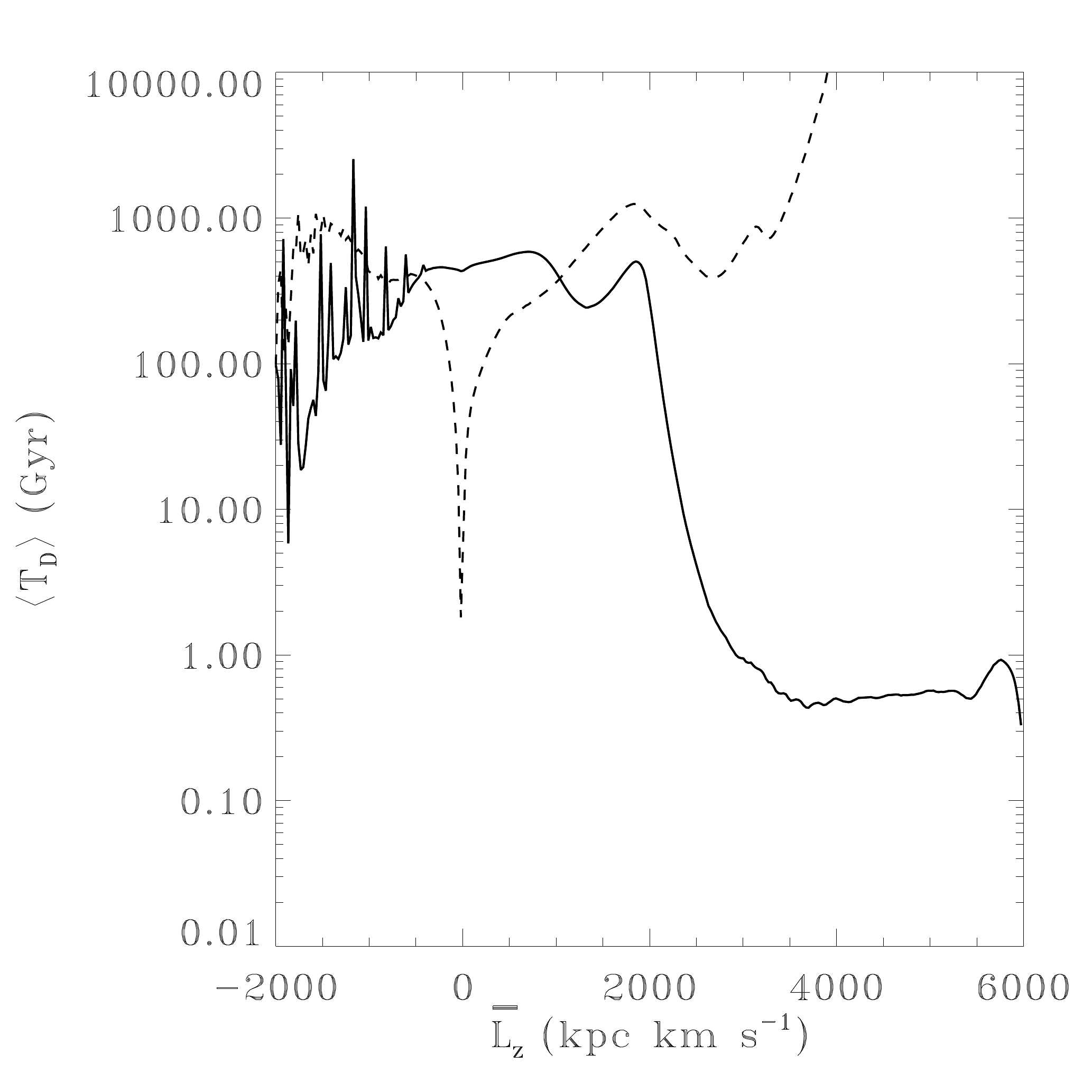}
  \caption{$\langle T_D(E)\rangle$ (full line) and $\langle
    T_D(L_\mathrm{z})\rangle$ (dashed line) as a function of
    ${\overline{L_\mathrm{z}}}$ for \BCPA, \FBFH, \PFAA\ and
    \PFAB\ from left to right. The shaded area delimits the region
    occupied by bar corotation during the 6 (\BCPA\ and \FBFH) and
    7~Gyr (\PFAA) evolution. Vertical lines show \Lz\ for the
    innermost {bar} UHR (dot-dashed) and the outermost OLR
    (long-dashed) positions reached.}
    \label{fig:td_lz}
\end{figure*}

In Figure~\ref{fig:td_lz} we have displayed diffusion time scales
$T_D(E)$ and $T_D(L_\mathrm{z})$ averaged over sets of particles
sampled by ${\overline{L_\mathrm{z}}}$ ranges, for the case of
\PFAA. ${\overline{L_\mathrm{z}}}$ is now time-averaged over $\approx
7$~Gyr. It is here a proxy for the mean radial position of
particles. In this $\langle T_D \rangle - {\overline{L_\mathrm{z}}}$
plot, we can overlay the approximate position of bar resonances
determined in the linear epicyclic approximation. Since the entire
resonance system moves outwards during the evolution of the galaxy, we
have plotted the position of the UHR at $t=3.16$~Gyr and the OLR at
$t=10.54$~Gyr. This delimits the range in \Lz\ occupied by all bar
resonances during the evolution. The specific range covered by the
corotation is approximately represented by the shaded area but can be
more extended when \rlun\ and \rlquatre\ positions are considered.

This averaged view of time scales and angular momentum confirms the
statements made in Section~\ref{sec:timescales}. In average, $\langle
T_D(E)\rangle$ decreases from the centre to the outermost regions. In
the range of ${\overline{L_\mathrm{z}}}$ delimited by resonances
positions, $\langle T_D(E)\rangle$ remains below 1~Gyr, and goes down
to 0.1~Gyr. Typical $\langle T_D(E)\rangle$ values outside OLR remain
of the order of the Gyr or below.

In contrast to $\langle T_D(E)\rangle$, $\langle T_D(Lz)\rangle$
increases from the centre outwards. In the bar resonance region, it
reaches values much higher than 10~Gyr.  The
diffusion of \Lz\ is therefore more effective well inside the UHR. In
order to avoid any misunderstanding, we remind here that we only study
the phase after the formation of the bar ($t> 3$~Gyr), in a regime
that can be considered as quiet.

There have been some debates on the action of the OLR as a
  barrier to stellar migration \citep{2015A&A...578A..58H,
    2016MNRAS.461.3835M}. In the case of our simulations, $\langle
  T_D(E)\rangle$ does indeed show a bump just after the OLR but the
  characteristic time scale remains of the Gyr order. $\langle
  T_D(Lz)\rangle$ continues to increase well beyond the bar OLR, even
  for \PFAB. Therefore, we can not confirm a specific signature of a
  barrier due to the bar OLR.

\begin{figure}[htb!]
  \centering
  \includegraphics[keepaspectratio,width=\hsize]{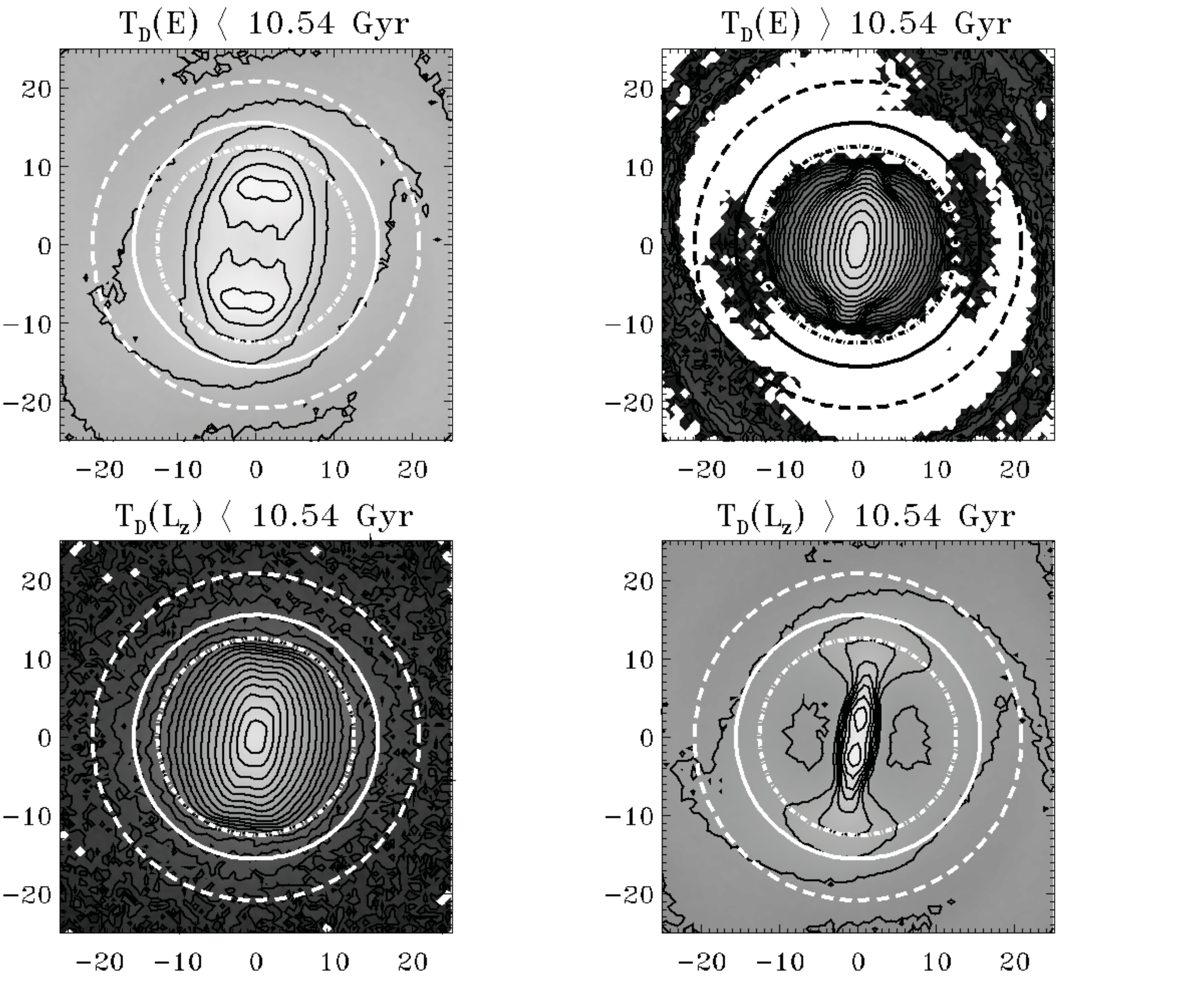}
  \caption{Projection of mass distribution on $x-y$ plane for
    particles selected according to $T_D(E)$ (top) and
    $T_D(L_\mathrm{z})$ (bottom) for \PFAA\ at $t=10.54$~Gyr. The
    $\log$ greyscale is identical for all four figures. Contour labels
    are spaced by 0.25 dex. The white or black circles show the
    position for the innermost UHR (dot-dashed), the corotation (full
    line), and the outermost OLR (long-dashed). Spatial scale is in
    kpc.}
    \label{fig:td_xy}
\end{figure}

Formally, \Lz\ is a good proxy for the radius only for orbits close to
circular. In order to verify the true spatial distribution, the mass
distribution obtained for short and long diffusion times (arbitrarily
defined as shorter or longer than the simulation length) can be
projected. Figure~\ref{fig:td_xy} shows the mass projected in the
x$-$y plane for $T_D$ less than or greater than 10.54~Gyr. It globally
confirms the analysis of Sect.~\ref{sec:timescales} but suggests that
a more detailed analysis must be performed. A few features deserve to
be mentioned. Particles with $T_D(E) > 10.54$~Gyr includes a
population that might be trapped around $L_{4,5}$ Lagrangian
points. They also included stable $x_1$ orbits inside the UHR, where
their shape is purely elliptical
\citep[][]{1983A&A...117...89C}. Beyond the UHR bifurcation, $x_1$
orbits become rectangular-like, develop loops, and can become
unstable. Their contribution is visible in the mass distribution for
$T_D(E) < 10.54$~Gyr. A much more detailed study of the $T_D$ spatial
distribution, and its relation to orbits families, is postponed to a
dedicated future paper.

\subsection{Stochastic diffusion}

Results similar to \citet[][]{2011A&A...534A..75B} have been reached:
the role of the `hot' population is highlighted in both studies and
the diffusion time scale depends on the radial position.  However, a
quantitative comparison with \citet[][]{2011A&A...534A..75B} is not
straightforward. The definition of their diffusion coefficient is
different and they expressed it as a function of time and radius.
Their study is based on Fourier's law of heat conduction. Heat
conduction is a non-equilibrium phenomenon.  A coarse-grained
description of the phenomenon with a clear separation between
microscopic and macroscopic scales can be assumed. At the microscopic
scale, heat carriers which are molecules and atoms in gas and liquids,
phonons in solids, evolve as a result of a deterministic Hamiltonian
description, whereas at macroscopic scale phenomenological Fourier's
law implies a diffusive transfer of energy.  However a rigorous
derivation of this law starting from a microscopic Hamiltonian
description is still lacking \citep[][]{2008AdPhy..57..457D}.

Microscopically we have to think about heat carriers colliding
randomly and the result is a heat diffusion.  However, in a pure
stellar N$-$body system, hard collisions are rare. Encounters are the
dominant process, especially weak ones, which makes the Fokker-Planck
equation the traditional tool for the study of stellar systems through
the frictional and diffusion coefficients
\citep[][]{1992rcd..book.....L, 2008gady.book.....B}. 

It can be shown that a DF whose evolution over time is governed by the
Fokker-Planck equation also follows a diffusion equation of Fourier's
form, with the same diffusion coefficient, provided that a
relationship with the friction coefficient is respected
\citep[cf][]{1992rcd..book.....L}. Therefore, the formalism used by
\citet[][]{2011A&A...534A..75B} may be similar to that of
Fokker-Planck.  

For simple Hamiltonians, a generalised Fokker-Planck
equation can be derived to include the energy drift due to scattering
and fast transport in phase space due to trapping/escape. This
derivation goes beyond the purpose of this paper.

The Chirikov coefficient implicitly includes all effects due to
resonances, and resonances overlaps due to several forcing patterns,
as well as effects due to a noisy potential.

\subsection{Limitations}

For our first paper on this topic, we have decided to restrict the
exploration of $E$ and \Lz\ variations to the simplest type of
simulations, the pure N$-$body case. Indeed, the absence of a gaseous
component is a major main limitation that has several clear
consequences. Without gas, there is no possibility to form a new
population whose kinematics might cool down the disc \citep[][for
  instance]{2015A&A...575A...7W}.  Another missing fluid is dark
matter. The main effect of a live dark halo (except to flatten the
rotation curve of the disc at a large distance) is to permit the
exchange of angular momentum with the stellar disc. The rate and the
amplitude of these exchanges depend on the velocity dispersion of both
the disc and the halo, and on the relative halo mass.
Depending on the rate at which the stellar disc losses its angular
momentum, the bar grows quite differently. Considering
\citet{2006ApJ...637..214M} simulations as representative, roughly 2/3
of the angular momentum loses by the bar-unstable part of the stellar
disc is absorbed by the halo, the rest going to the outer disc. Most
of these exchanges happen during the buckling of the bar, which, in
the case of our simulations, occurs for $t < 3$~Gyr. 

A final limitation comes from the genuine nature of galaxies, which
are much more complex than these idealised simulations. Much of this
complexity comes from perturbations by random sources. These sources
can be intrinsic (such as molecular clouds or Poissonian shot noise)
or extrinsic (satellites accretion, globular clusters, etc.). All
these perturbations could contribute to reduce diffusion times, but
this remains to be demonstrated in the specific case at hand.

\section{Conclusions}

We have computed \citet[][]{1979PhR....52..263C} diffusion rates
(\dem\ and \dlzm) and related diffusion time scales ($T_D(E)$ and
$T_D(L_\mathrm{z})$) for energy ($E$) and angular momentum (\Lz) in
pure N$-$body simulations of disc galaxies developing bars and spiral
structures. These quantities were only calculated once the bar was
perfectly settled in order to reflect the evolution of the disc under
the effect of its presence.

We can summarise our results as follows:

\begin{enumerate}
  \item Both $E$ and \Lz\ diffuse during the evolution of a stellar
    disc, under the effect of intrinsic perturbations caused by the
    bar and spiral structures. In particular, bars and spiral
    structures are responsible for shortening diffusion time scales.
  \item Diffusion time scales are shorter than the simulations length
    (i.e. $\approx 10$~Gyr) for different particle populations
    depending on whether the diffusion of $E$ or \Lz\ is
    considered. Consequently, the regions affected by the diffusion
    differ according to the quantity that diffuses.
  \item The distribution function of Chirikov diffusion rates $D_2$
    has the same shape regardless the simulation considered. It can be
    approximate by the equation $ \log n(D_2)\propto \gamma D_2 /
    \max(D_2) $ where $\gamma\approx -5.0$ for \dem\ and $\gamma$ is
    in the range $[-6.61 ; -6.21]$ for \dlzm.
  \item At first order, values of \dem\ remain within the same range
    in axisymmetric and non-axisymmetric simulations unlike for \dlzm.
  \item $T_D(E)$ is shorter than simulation length for particles
    belonging to the `hot' population, the disc, and families of
    orbits lying between the bar UHR and corotation. It is minimal
    (and shorter than 1~Gyr) in the region delimited by the set of bar
    resonances (between UHR and OLR).
  \item $T_D(L_\mathrm{z})$ is shorter than simulation length mainly for
    particles inside the bar region (i.e. inside the UHR).
  \item On average, $T_D(L_\mathrm{z})$ increases with radius while
    $T_D(E)$ tends to decrease from the centre to the most external
    regions. 
  \item The so-called 'hot' population, which navigates between the
    bar and the disc, plays only an important role in diffusion of
    $E$.
\end{enumerate}

This article is limited to a first exploration of the results obtained
with the Chirikov diffusion rate. Next articles will explore in
greater depth the phenomena of migration, diffusion and resonance,
particularly in terms of orbital structure.

\acknowledgements
{First of all, I would like to thank the referee for her/his very
constructive comments. I would like to acknowledge the University of
Strasbourg HPC department and the Meso@LR computing centre of the
University of Montpellier for providing access to computing
resources. Part of the computing resources were funded by grants from
the IDEX Unistra and the Scientific Council of the University of
Montpellier.}

\bibliography{diffusion} 
\bibliographystyle{aasjournal} 

\appendix

\begin{figure*}[htb!]
  \centering
  \includegraphics[keepaspectratio,width=0.45\hsize]{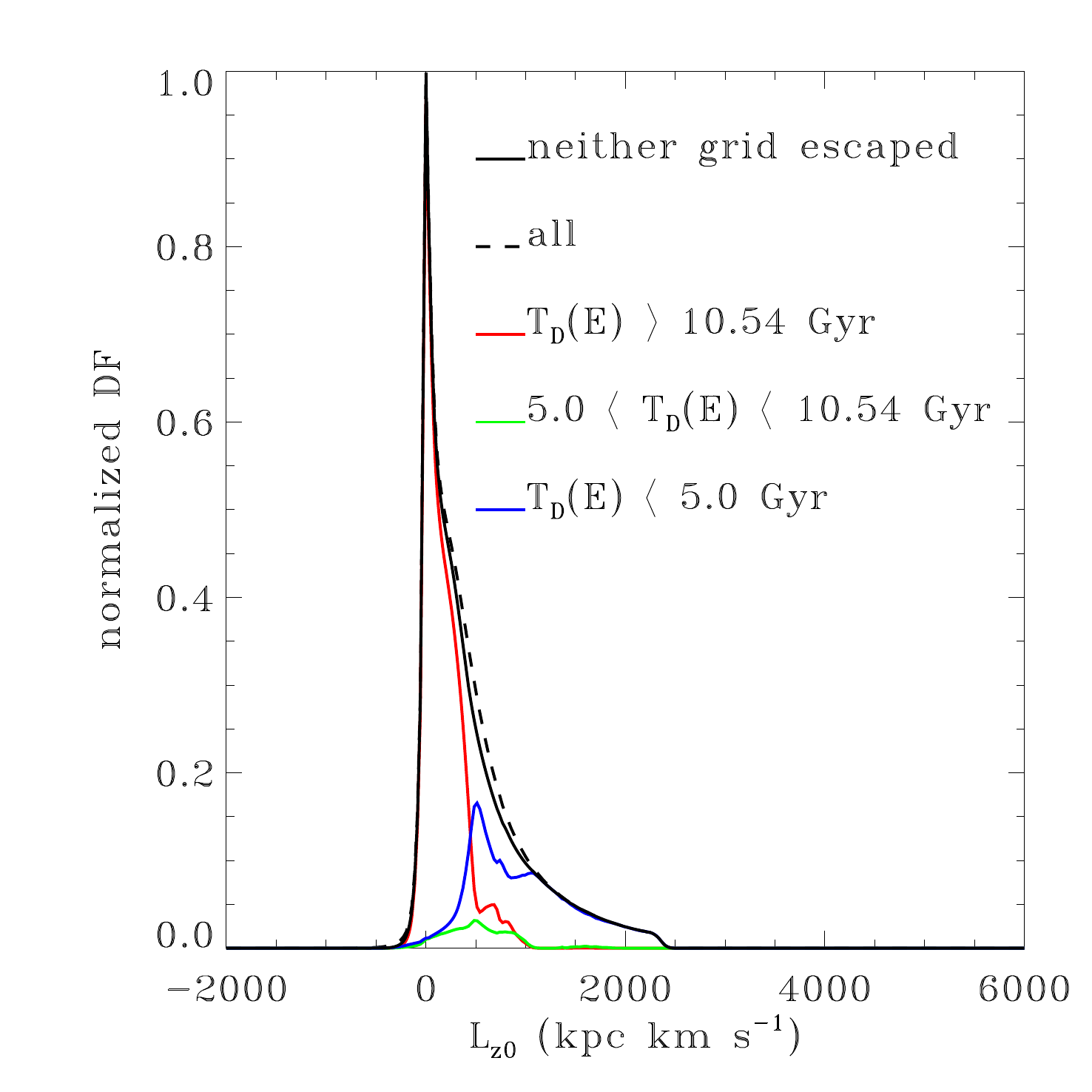}
  \includegraphics[keepaspectratio,width=0.45\hsize]{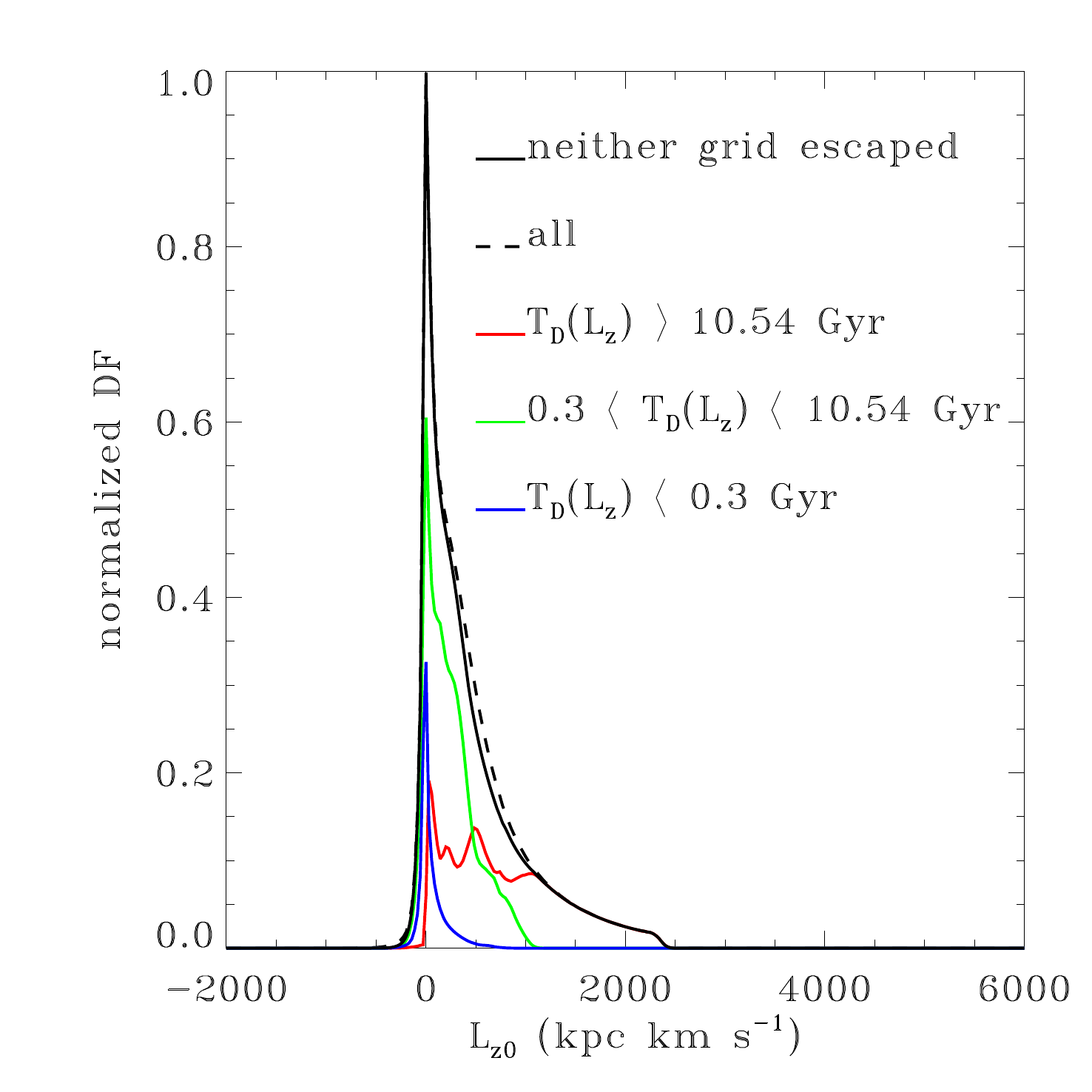}
  \caption{Top row: distribution function DF(\Lz) for \BCPA\ at $t=0$. Dashed
    line: all particles ($N_\mathrm{s}$). Solid line: only particles
    that neither escaped from the grid and used for $T_D(E)$
    computations. Red line: particles with $T_D(E) > 10.54$~Gyr. Green
    line: $T_D(E) > 4.0$~Gyr. Blue line: $T_D(E) < 4.0$~Gyr. All DFs
    have been normalised to DF(0) with all particles, which is also
    the maximum. Bottom row : same as top row but for
    $T_D(L_\mathrm{z})$. The noticeable time scale is now 0.3~Gyr
    instead of 4~Gyr.}
\end{figure*}
\begin{figure*}[htb!]
  \centering
  \includegraphics[keepaspectratio,width=0.45\hsize]{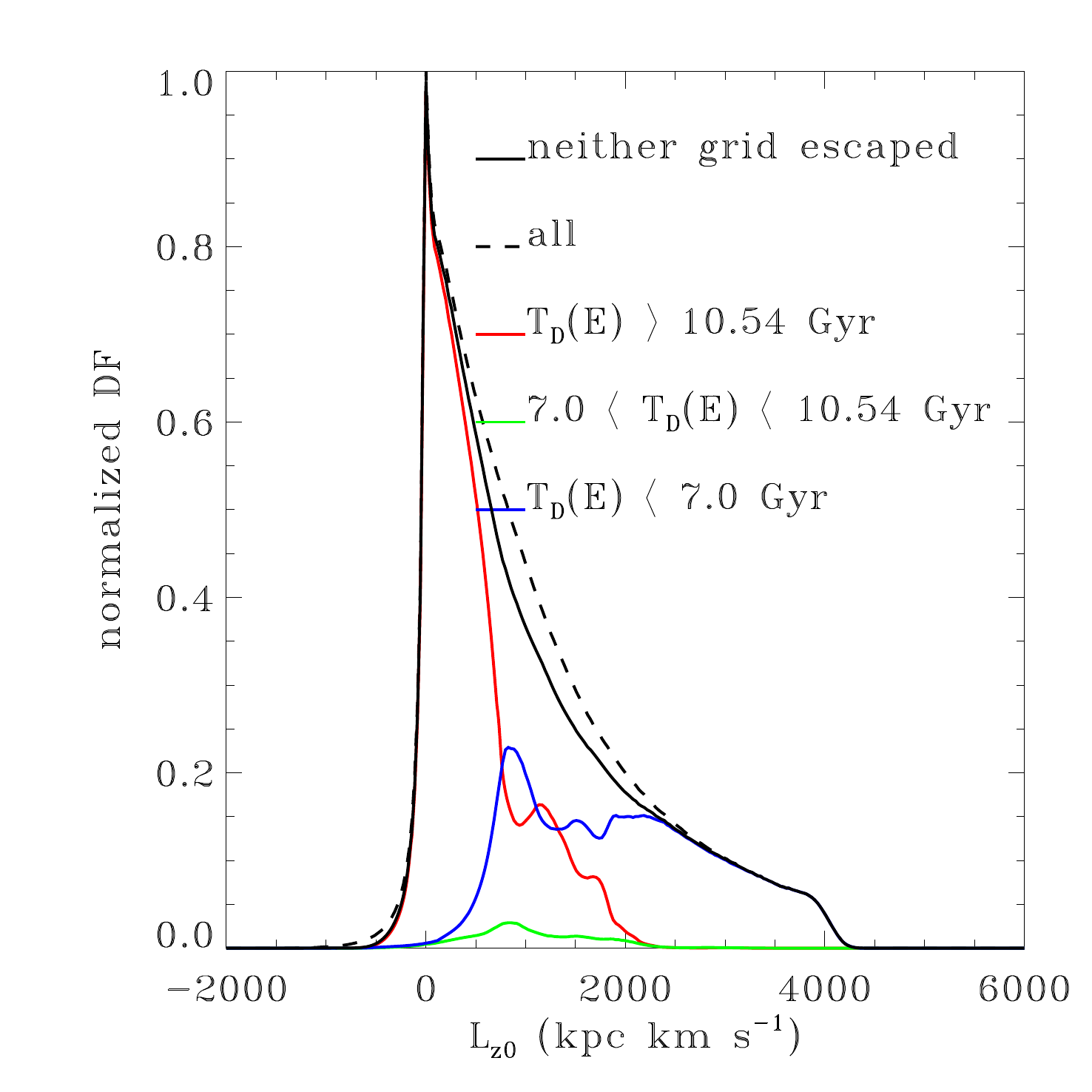}
  \includegraphics[keepaspectratio,width=0.45\hsize]{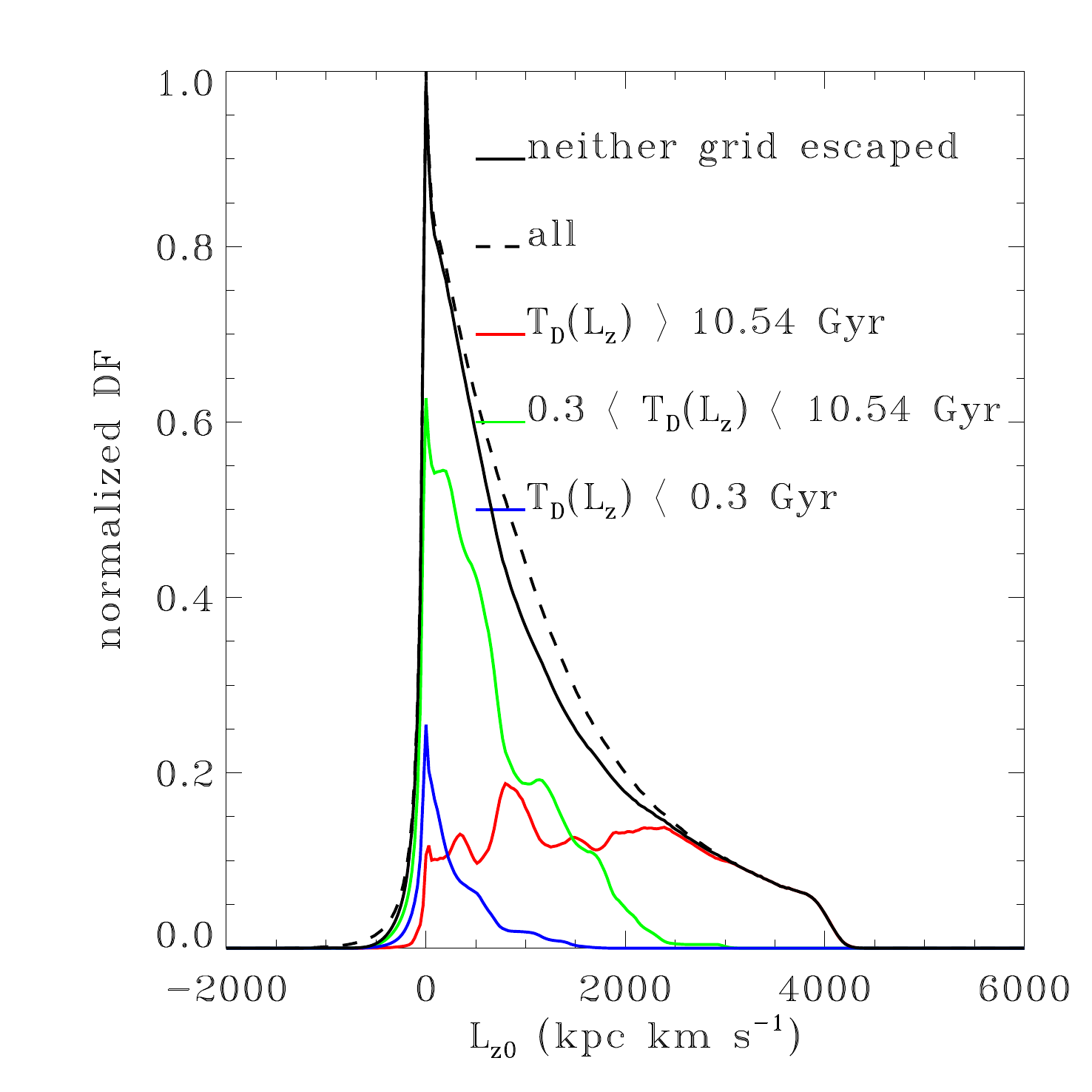}
  \caption{Top row: distribution function DF(\Lz) for \FBFH\ at $t=0$. Dashed
    line: all particles ($N_\mathrm{s}$). Solid line: only particles
    that neither escaped from the grid and used for $T_D(E)$
    computations. Red line: particles with $T_D(E) > 10.54$~Gyr. Green
    line: $T_D(E) > 4.0$~Gyr. Blue line: $T_D(E) < 4.0$~Gyr. All DFs
    have been normalised to DF(0) with all particles, which is also
    the maximum. Bottom row : same as top row but for
    $T_D(L_\mathrm{z})$. The noticeable time scale is now 0.3~Gyr
    instead of 4~Gyr.}
\end{figure*}
\end{document}